\documentclass[]{siamart1116}

\usepackage{amsfonts}
\usepackage{graphicx}
\usepackage{epstopdf}
\usepackage{algorithmic}
\ifpdf
  \DeclareGraphicsExtensions{.eps,.pdf,.png,.jpg}
\else
  \DeclareGraphicsExtensions{.eps}
\fi

\numberwithin{theorem}{section}

\usepackage{graphicx}
\usepackage{float}
\usepackage{subfigure}
\usepackage{wrapfig}
\usepackage{amscd}

\newcommand{\pext}{pdf}

\usepackage{hyperref}
\usepackage{enumitem}

\newcommand{\IE}{\mathbb{E}}
\newcommand{\IV}{\mathbb{V}}
\newcommand{\EMLMC}{E^{\text{MLMC}}}
\newcommand{\EOFMLMC}{E^{\text{OF-MLMC}}}

\newcommand{\IR}{\mathbb{R}}

\newcommand{\cB}{{\mathcal B}}

\newcommand{\cO}{{\mathcal O}}

\newcommand{\om}{\omega}
\newcommand{\Om}{\Omega}

\newcommand{\bQ}{{\mathbf Q}}

\newcommand{\bc}{{\mathbf c}}
\newcommand{\bx}{{\mathbf x}}

\newcommand{\bvec}[1]{{\mathbf #1}}
\newcommand{\ve}[1]{{\mathbf #1}}

\newcommand{\IP}{\mathbb{P}}

\newcommand{\Cov}{\operatorname{Cov}}

\newcommand{\Work}{\mathbb{W}}

\newcommand{\TheTitle}{Optimal fidelity multi-level Monte Carlo for quantification of uncertainty in simulations of cloud cavitation collapse}
\newcommand{\TheAuthors}{J. {\v S}ukys, U. Rasthofer, F. Wermelinger, P. Hadjidoukas, and P. Koumoutsakos}

\newcommand{\RunningTitle}{OF-MLMC for uncertainty quantification in cloud cavitation}
\newcommand{\RunningAuthors}{{\v S}ukys, Rasthofer, Wermelinger, Hadjidoukas, and Koumoutsakos}

\headers{\RunningTitle}{\RunningAuthors}

\title{{\TheTitle}\thanks{Submitted to the editors 9 May 2017.
\funding{INCITE project CloudPredict; Argonne Leadership Computing Facility, DOE project DE-AC02-06CH11357; PRACE: JFZ and CINECA, projects PRA091 and Pra09\_2376;
CSCS project ID s500.}}}

\author{
  Jonas {\v S}ukys\thanks{Computational Science and Engineering Laboratory, ETH Z\"urich, Switzerland. Current address: Eawag, Swiss Federal Institute of Aquatic Science and Technology, Switzerland.
    (\email{jonas.sukys@eawag.ch}, \url{science.sukys.ch}).}
  \and
  Ursula Rasthofer\thanks{Computational Science and Engineering Laboratory, ETH Z\"urich, Switzerland
    (\email{urasthofer,fabianw,phadjido,petros@mavt.ethz.ch}, \url{cse-lab.ethz.ch}).}
  \and
  Fabian Wermelinger\footnotemark[3]
  \and
  Panagiotis Hadjidoukas\footnotemark[3]
  \and
  Petros Koumoutsakos\footnotemark[3]\hspace*{4pt}\thanks{Corresponding author.}
}

\usepackage{amsopn}

\ifpdf
\hypersetup{
  pdftitle={\TheTitle},
  pdfauthor={\TheAuthors}
}
\fi

\begin{document}

\maketitle

\begin{abstract}
We quantify uncertainties in the location and magnitude of extreme pressure spots revealed from large scale multi-phase flow simulations of cloud cavitation collapse.
We examine clouds containing 500 cavities and quantify uncertainties related to their  initial spatial arrangement. The resulting 2000-dimensional space is sampled using a non-intrusive and computationally efficient Multi-Level Monte Carlo (MLMC) methodology.
We introduce novel optimal control variate coefficients to enhance the variance reduction in MLMC. The proposed optimal fidelity MLMC leads to more than two orders of magnitude speedup when compared to standard Monte Carlo methods.
We identify large uncertainties in the location and magnitude of the peak pressure pulse and present its statistical correlations and joint probability density functions with the geometrical characteristics of the cloud.
Characteristic properties of spatial cloud structure are identified as potential causes of significant uncertainties in exerted collapse pressures.
\end{abstract}

\begin{keywords}
  compressible multi-phase flow,
  high performance computing,
  diffuse interface method,
  bubble collapse,
  cloud cavitation,
  uncertainty quantification,
  multi-level Monte Carlo,
  optimal control variates,
  fault tolerance
\end{keywords}

\begin{AMS}
  76B10, 
  68W10, 
  65C05, 
  65C60, 
  68M15. 
\end{AMS}

\section{Introduction}

\begin{figure}[ht]
	\centering
	\includegraphics[width=0.48\textwidth]{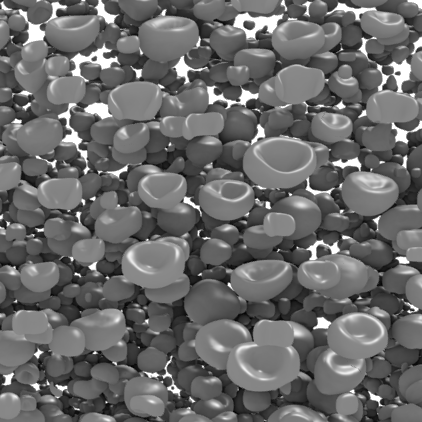}
    \hfill
	\includegraphics[width=0.48\textwidth]{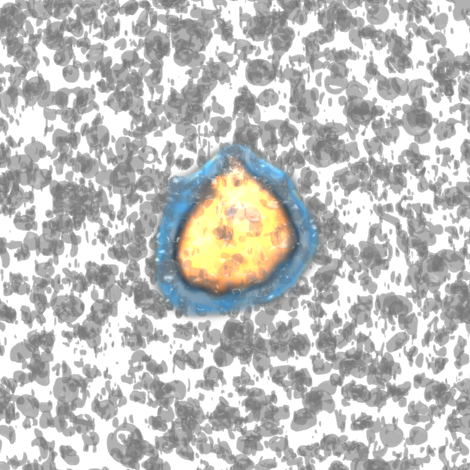}
	\caption[Collapsing cloud containing 50'000 gas cavities]
	{Surface view of a collapsing cloud containing 50'000 gas cavities (left)
    and pressure peak at the center of the cloud (right).
	Outer bubbles evolve into cap-like shapes, forming an array of inwards directed velocity micro-jets.
    The resulting focused pressure peak amplifications of two orders of magnitude are exerted at the cloud center \cite{hadjidoukas2015b}.}
	\label{f:cloud-50K}
\end{figure}

Cloud cavitation collapse pertains to the inception of multiple gas cavities in a liquid,
and their subsequent rapid collapse driven by an increase of the ambient pressure.
Shock waves emanate from the cavities with pressure peaks up to two orders of magnitude larger than the ambient pressure \cite{Morch:1980,Chahine:1984,Brennen:1995}.
When such shock waves impact on solid walls, they may cause material erosion,
considerably reducing the performance and longevity of turbo-machinery and fuel injection engines \cite{Schmidt:2001,Li:2015}.
On the other hand, the destructive potential of cavitation can be harnessed for non-invasive biomedical applications \cite{Coussios:2008,brennen2015a}
and efficient travel in military sub-surface applications \cite{Bonfiglio2017}.
Prevalent configurations of cavitation often involve clouds of hundreds or thousands of bubbles \cite{Brennen:1995}.
The cloud collapse entails a location dependent, often non-spherical, collapse of each cavity
that progresses from the surface to the center of the cloud.
Pressure waves emanating from collapsing cavities located near the surface of the cloud
act as amplifiers to the subsequent collapses at the center of the cloud.
The interaction of these pressure waves increases the destructive potential as compared to the single bubble case.
Cavitation, in particular as it occurs in realistic conditions, presents a formidable challenge to
experimental \cite{bremond2006a,brujan2012a,yamamoto2016a} and computational \cite{Adams:2013,tiwari2015a} studies
due to its geometric complexity and multitude of spatio-temporal scales. 
Blake et al.~\cite{Blake:1997} studied the single bubble asymmetric collapse using a boundary integral method.
Johnsen and Colonius \cite{Johnsen:2009} investigated the potential damage of single collapsing bubbles
in both spherical and asymmetric regime for a range of pulse peak pressures in shock-induced collapse.
Lauer et al.~\cite{Lauer:2012} studied collapses of arrays of cavities under shock waves
using the sharp interface technique of Hu et al.~\cite{HuAdams:2006}.
Recent numerical investigation of cloud cavitation involved a cluster of 125 vapor bubbles
inside a pressurized liquid at 40 bar \cite{Schmidt:2011,Adams:2013},
and a cluster of 2500 gas bubbles with ambient liquid pressure of 100 bar \cite{iccs}.
Large scale numerical simulations of cloud cavitation collapse
considered clouds containing 50'000 bubbles \cite{hadjidoukas2015b}. 
Visualizations of such a collapsing cloud and the resulting focused pressure peak
at the center are reproduced in \autoref{f:cloud-50K}.
However the computational demands of these simulations do not allow for further parametric studies.

\begin{figure*}[bht]
	\begin{center}
		\begin{tabular}{c}
			\includegraphics[width=0.42\textwidth]{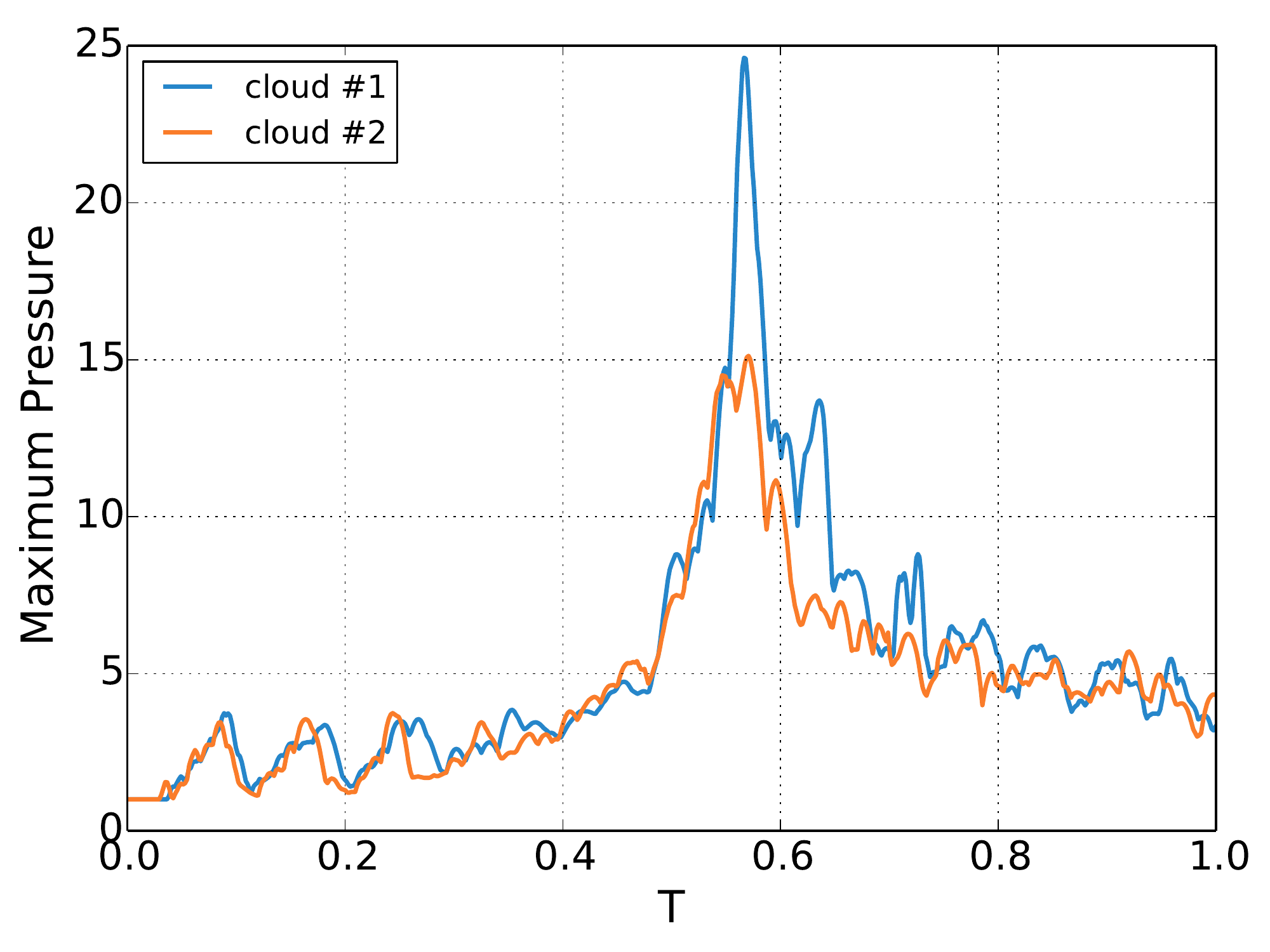}
		\end{tabular}
	\end{center}
	\caption{
		Evolution of the normalized peak pressure versus normalized time
		in free-field collapse of two spherical clouds containing 1'000 bubbles. 
		Two samples are drawn from the same uniform distribution (for the positions of the cavities)
		and log-normal distribution (for the radii of the cavities).
		\label{fig:two_sampels}}
\end{figure*}

A challenge in modeling and quantifying cloud cavitation collapse is the dependence of critical 
Quantities of Interest (QoIs), such as peak pressure or collapse time,
on a particular (random) cloud configuration \cite{tiwari2015a} (see also Figure \ref{fig:two_sampels}).
The systematic study of such dependencies can be addressed through an Uncertainty Quantification (UQ) framework,
recently applied in \cite{Congedo,Bonfiglio2017}. 
In \cite{schwabsid1,MSS11-syscl,MSS11-sw}, a mathematical framework was provided for uncertain solutions of hyperbolic equations.
Popular probability space discretization methods include generalized Polynomial Chaos (gPC) techniques
(please see \cite{CGH1,LSK1,TLNE,PDL,WK1,GX08} and references therein).
An alternative class of methods for quantifying uncertainty in PDEs are the stochastic collocation methods \cite{XH1,MZ1,WLB1}.
However, the lack of regularity of the solution with respect to the stochastic variables
impedes efficient performance of both the stochastic Galerkin as well as the stochastic collocation methods,
in particular for high-dimensional parameter spaces.
Here, we propose the development and implementation of non-intrusive
Monte Carlo (MC) methods for UQ of cloud cavitation collapse.
In MC methods, the governing  equations are  solved for a sequence of randomly generated samples, which
are combined to ascertain statistical information.
However, the robustness of MC methods with respect to solution regularity
comes at the price of a low error convergence rate  regarding the number of samples.
Drawbacks of the mentioned numerical uncertainty quantification methods inspired the development of various multi-fidelity methods, such as Multi-Level Monte Carlo (MLMC) \cite{GIL2},
multi-fidelity recursive co-kriging and Gaussian--Markov random field \cite{Soc2015}.
Further developments include multi-fidelity
Gaussian Process Regression (GPR) based on co-kriging \cite{Parussini2017},
and purely data-driven algorithms for linear equations using Gaussian
process priors to completely circumvent the use of numerical discretization schemes \cite{Raissi2017}.
MLMC methods were introduced by Heinrich for numerical quadrature \cite{H1},
then pioneered by Giles for It\^{o} SPDEs \cite{GIL2},
and have been lately applied to various stochastic PDEs \cite{BScZ10,CGST1,GR1,MJM12}.
The MLMC algorithm was also extended to hyperbolic conservation laws and to massively parallel simulations
of the random multi-dimensional Euler, magneto-hydrodynamics (MHD) and shallow water equations \cite{schwabsid1,MSS11-syscl,MSS11-sw,MSS14-wave,SMS11-static,Sukys13-adaptive}.
Subsequent MLMC improvements include Bayesian inference
for fusing knowledge on empirical statistical estimates and deterministic convergence rates \cite{Tempone},
an alternative multi-fidelity setting where sample recycling with optimal control variate coefficients are employed \cite{Gunzburger,Peherstorfer2016},
and multi-level Markov Chain Monte Carlo posterior sampling in Bayesian inference problems \cite{Dodwell2015AFlow,Elsakout2015MultilevelQuantification}.

The ensemble uncertainty of the clouds is parametrized by means of probability distributions
of cavity radii, positions and initial internal pressures.
Our goal is to perform simulations of cloud cavitation collapse with unprecedented number of interacting cavities
with full UQ on QoIs (peak pressure magnitudes, locations and cloud collapse times)
in terms of means, variances, confidence intervals and even probability density functions.
To the best of our knowledge, such extensive UQ analysis of (even small) clouds has not been accomplished before.
We propose an improved
non-intrusive MLMC method
with novel optimal control variate coefficients
coupled with the state-of-the-art numerical solver for
cloud cavitation collapse simulations and
provide robust statistical estimates on relevant quantities of interest,
emphasizing the importance of UQ in such problems.

The paper is structured as follows:
\autoref{s:numerics} introduces the governing multi-phase equations and the finite volume method used for their solution. It also presents the optimal control variate MLMC method for the statistical sampling of QoIs.
Numerical experiments quantifying such uncertainties and identifying their relations to the geometric properties of the cloud by means of joint probability density function estimates are reported in \autoref{s:results}.
We summarize our findings in \autoref{s:summary}.

\section{Computational methods}
\label{s:numerics}

The governing system of equations for multi-phase flows are discretized using a  Finite Volume Method (FVM) that is efficiently implemented so as to take advantage of supercomputer architectures.
The sampling needed to estimate statistical properties from an ensemble of evolved QoIs is performed  by a novel  Optimal Fidelity MLMC (OF-MLMC) method. We introduce a Python implementation PyMLMC \cite{pymlmc} of OF-MLMC and its embedding in an open source uncertainty quantification framework \cite{Pi4U}.

\subsection{Governing equations}
\label{ss:equations}

The dynamics of cavitating and collapsing clouds of bubbles are governed by the compressibility of the flow, with viscous dissipation and capillary effects taking place at orders of magnitude slower time scales. Hence, we describe cloud cavitation collapse through the Euler equations for inviscid, compressible, multi-phase flows.
The system of equations, derived from the Baer-Nunziato model~\cite{baer1986a}, describes the  
evolution of density, momentum and total energy of the multi-phase flow in the domain $D \subset \IR^3$ as \cite{Murrone:2005,Perigaud:2005}:
\begin{align}
\label{eq:mass_1}
\frac{\partial \alpha_1\rho_1}{\partial t} 
+ \nabla\cdot\left(\alpha_1\rho_1 \ve u \right) & = 0,\\
\label{eq:mass_2}
\frac{\partial \alpha_2\rho_2}{\partial t} 
+ \nabla\cdot\left(\alpha_2\rho_2 \ve u \right) & = 0,\\
\label{eq:momentum}
\frac{\partial \left(\rho \ve u \right)}{\partial t} + \nabla\cdot 
\left(\rho \ve u \otimes \ve u + p \ve I \right) & = \ve 0,\\
\label{eq:energy}
\frac{\partial E}{\partial t} + \nabla\cdot\left( \left(E+p \right)\ve 
u \right) & = 0,\displaybreak[0]\\
\label{eq:vol_frac}
\frac{\partial \alpha_2}{\partial t} + \ve u\cdot\nabla\alpha_2  
& = K\,\nabla\cdot\ve u,
\end{align}
where
\begin{equation}
\label{eq:var_K}
K=\frac{\alpha_1\alpha_2(\rho_1 c_1^2 - \rho_2 c_2^2 )}{\alpha_1\rho_2 
	c_2^2 + \alpha_2\rho_1 c_1^2}.
\end{equation}
All quantities, unless otherwise stated, depend on spatial variable $\bx \in D$ and time variable $t \geq 0$.
This  system comprises two mass conservation equations, one for each 
phase, conservation equations for momentum and total energy in single- 
or \mbox{mixture-fluid} formulation as well as a transport equation for the volume 
fraction of one of the two phases with source/sink term on the right-hand 
side. In~\eqref{eq:mass_1}--\eqref{eq:vol_frac}, $\ve u$ denotes the 
velocity, $p$ the pressure, $\ve I$ the identity tensor, $\rho$ the (mixture) density, $E$ the (mixture) 
total energy $E=\rho e + 1/2 \rho (\ve u \cdot \ve u)$, where $e$ is the 
(mixture) specific internal energy.  Moreover, $\rho_k$, $\alpha_k$ and $c_k$ 
with  $k \in \lbrace 1,2 \rbrace$ are density, volume fraction and speed of 
sound of the two phases. For the mixture quantities, the following additional relations hold: $\alpha_1 + \alpha_2=1$, 
$\rho = \alpha_1\rho_1 + \alpha_2\rho_2$, and $\rho e = \alpha_1\rho_1e_1 
+ \alpha_2\rho_2e_2$.
We do not account for  mass transfer between different phases (evaporation or condensation),
so that the above equations describe a multi-component flow.
The source term in \eqref{eq:vol_frac} for homogeneous 
mixtures \cite{Kapila:2001} describes the reduction of the gas volume fraction in a mixture of 
gas and liquid when a compression wave travels across the mixing region and vice versa for an expansion wave. For a more detailed analysis on the positive influence of this term on the accuracy of the model equations, we refer to \cite{iccs}. 

The equation system is closed by an appropriate equation of state for each of 
the phases. We employ the  stiffened equation of 
state (see \cite{Menikoff:1989} for a review) to capture liquids and gases. This  enables a simple, analytic approximation to arbitrary fluids and is expressed by
\begin{equation}
p = \left(\gamma_k -1\right)\rho_k e_k - \gamma_k p_{\mathrm{c},k},
\end{equation}
where isobaric closure is assumed \cite{Perigaud:2005}.
Parameters $\gamma_k$ and $ p_{\mathrm{c},k}$ depend on the material. 
For $ p_{\mathrm{c},k}=0$, the equation of state for ideal gases is recovered.
For simulations in this manuscript, $\gamma_1= 6.59$ and $p_\mathrm{c,1}= 4.049\cdot 10^{8}\,\mathrm{Pa}$
are used for water and $\gamma_2= 1.4$ and $p_\mathrm{c,2}= 0\,\mathrm{Pa}$ for air.

\subsection{Numerical method}
\label{ss:fvm}

The governing system \eqref{eq:mass_1}--\eqref{eq:vol_frac} can be recast 
into the quasi--conservative form
\begin{align}
\label{eq:system}
\frac{\partial \bvec{Q}}{\partial t} + \frac{\partial \bvec{F}(\bvec{Q})}{\partial x} +
\frac{\partial \bvec{G}(\bvec{Q})}{\partial y} + \frac{\partial \bvec{H}(\bvec{Q})}{\partial z}
&= \bvec{R}(\bvec{Q}),
\end{align}
where $\bvec{Q}=(\alpha_1\rho_1,\alpha_2\rho_2,\rho{\mathbf u},E,\alpha_2)^T$
is the vector of conserved (except for $\alpha_2$ which has a non-zero source term) variables
and $\bvec{F}(\bvec{Q})$, $\bvec{G}(\bvec{Q})$, $\bvec{H}(\bvec{Q})$ are vectors of flux functions
\begin{align*}
\bvec{F}(\bvec{Q})=
\begin{pmatrix}
\alpha_1\rho_1 u_x\\
\alpha_2\rho_2 u_x\\
\rho u_x^2 + p\\
\rho u_y u_x\\
\rho u_z u_x\\
(E+p)u_x\\
\alpha_2 u_x
\end{pmatrix}\!,\quad
\bvec{G}(\bvec{Q})=
\begin{pmatrix}
\alpha_1\rho_1 u_y\\
\alpha_2\rho_2 u_y\\
\rho u_x u_y\\
\rho u_y^2 + p\\
\rho u_z u_y\\
(E+p)u_y\\
\alpha_2 u_y
\end{pmatrix}\!,\quad
\bvec{H}(\bvec{Q})=
\begin{pmatrix}
\alpha_1\rho_1 u_z\\
\alpha_2\rho_2 u_z\\
\rho u_x u_z\\
\rho u_y u_z\\
\rho u_z^2 + p\\
(E+p)u_z\\
\alpha_2 u_z
\end{pmatrix}\!.
\end{align*}
The source term $\bvec{R}(\bvec{Q})$ has all elements equal to zero except the last one
\begin{equation*}
\bvec{R}(\bvec{Q})_7 = \alpha_2(\nabla\cdot{\mathbf u}) + K\,\nabla\cdot\ve u,
\end{equation*}
which appears due to rewriting~\eqref{eq:vol_frac} in conservative form~\cite{johnsen2006a} and incorporating the present source term.
For the system \eqref{eq:system}, initial condition $\bQ(\bx, t = 0) = \bQ_0(\bx)$  over the entire domain $\bx \in D$
as well as boundary conditions at $\bx \in \partial D$ for all times $t \geq 0$ need to be provided to complete the full specification of the multi-phase flow problem.
The method of lines is applied to 
obtain a semi-discrete representation 
of~\eqref{eq:system}, where space 
continuous operators are approximated using the FVM for uniform  structured grids.
The approach yields a system of ordinary differential equations
\begin{align}
\label{eq:ode}
\frac{d\ve{V}(t)}{dt} &= \mathcal{L}(\ve{V}(t)),
\end{align}
where $\ve{V}$ is a vector of cell average values and $\mathcal{L}(\cdot)$ is 
a discrete operator that approximates the convective fluxes and the given sources in the governing 
system.  The temporal discretization of~\eqref{eq:ode} is obtained by an explicit 
third-order low-storage Runge-Kutta 
scheme \cite{williamson1980a}.  The computation of the numerical fluxes is 
based on a Godunov-type scheme using the approximate HLLC Riemann 
solver originally introduced for single-phase flow problems in~\cite{toro1994a}.  The Riemann initial states are determined by 
a shock capturing third- or fifth-order accurate WENO reconstruction (see~\cite{jiang1996a}).  
Following~\cite{johnsen2006a}, the reconstruction is carried out using primitive variables, and
the HLLC Riemann solver is adapted to~\eqref{eq:vol_frac}
to prevent oscillations at interface.   
The solution is advanced 
with a time--step size that satisfies the Courant-Friedrichs-Lewy (CFL) condition.
For the coefficient weights in the Runge-Kutta 
stages, the values suggested in~\cite{gottlieb1998a} are used, resulting in a total variation 
diminishing scheme.

\subsection{Cubism-MPCF}
\label{ss:cubism}

The FVM used for the spatio-temporal numerical discretization of non-linear system of conservation laws in \eqref{eq:system} is implemented in the open source software Cubism-MPCF \cite{rossinelli2013a,hadjidoukas2015a,hadjidoukas2015b,iccs}.
The applied scheme entails three computational kernels: computation of CFL-limited time-step size $\Delta t$ based on a global reduction, evaluation of the approximate Riemann problem corresponding to the evaluation of the right-hand side $\mathcal{L}$ in \eqref{eq:ode} for each time step, and appropriate Runge-Kutta update steps.
%
%
The solver is parallelized with a hybrid paradigm using the MPI and OpenMP programming models.
The software is split into three abstraction layers: cluster, node, and core \cite{hadjidoukas2015a}.
The realization of the Cartesian domain decomposition and the inter-rank MPI communication is accomplished on the cluster layer.
On the node layer, the thread level parallelism exploits
the OpenMP standard using dynamic work scheduling.
Spatial blocking techniques are used to increase locality of the data,
with intra-rank ghost cells obtained from loading fractions of the surrounding blocks,
and inter-rank ghost cells obtained from a global buffer provided by the cluster layer.
On the core layer, the actual computational kernels
are
executed, exploiting data level parallelism and instruction level parallelism,
which are enabled by the conversion between the array--of--structures and structure--of--arrays layout.
For the simulations reported here, main parts of the computations were executed in mixed precision arithmetic.
More details on software design regarding the parallelization and optimization strategies used in Cubism-MPCF can be found in \cite{hadjidoukas2015a,hadjidoukas2015b,rossinelli2013a,iccs}.
Cubism-MPCF has demonstrated state--of--the--art performance in terms of floating point operations, memory traffic and storage, exhibiting  almost perfect overlap of communication and computation \cite{hadjidoukas2015a,iccs}. The software has been optimized to take advantage of the IBM BlueGene/Q (BGQ) and Cray XC30 platforms to simulate cavitation collapse dynamics using up to 13~Trillion computational cells with very efficient strong and weak scaling up to the full size of MIRA (Argonne National Laboratory) and Piz Daint (Swiss Supercomputing Center) supercomputers \cite{rossinelli2013a,hadjidoukas2015b,iccs}.

\subsection{Multi-Level Monte Carlo (MLMC) method}

In this section, we introduce the MLMC  framework for UQ. We  also present a novel and improved numerical sampling method for approximating the  statistical QoI.

This study is grounded on the theoretical probabilistic framework for non-linear systems of conservation laws introduced in \cite{MSS11-syscl,Sukys14-dissertation}.
Uncertainties in the system of conservation laws \eqref{eq:system},
such as uncertain initial data at time $t = 0$ for the vector of conserved quantities $\bQ$,
are modeled as \emph{random fields} \cite{MSS11-syscl,Sukys14-dissertation}. They 
depend  on the spatial and temporal variables $\bx$ and $t$,
as well as  on the stochastic parameter $\om \in \Om$,
which represents variability in the cloud configuration.
For instance, for uncertain initial data, we assume
\begin{equation}
\bQ_0(\bx,\om) = \bQ(\bx,0,\om) \in \IR^7,
\qquad \bx \in D, \quad \om \in \Om,
\end{equation}
i.e., at every spatial coordinate $\bx$, initial data $\bQ_0 (\bx, \om)$ is a random vector containing $7$ values, one for each equation in \eqref{eq:system}.
We further assume that $\bQ_0$ is such that at every spatial point $\bx$ the statistical moments such as expectation and variance exist and are defined by
\begin{equation}
\IE[\bQ_0](\bx) = \int_\Om \bQ(\bx,\om) d \om
\end{equation}
and
\begin{equation}
\IV[\bQ_0](\bx) = \int_\Om \Big(\bQ_0(\bx,\om) - \IE[\bQ_0](\bx)\Big)^2 d \om.
\end{equation}
Such uncertainties, for instance in initial data $\bQ_0$, are propagated according to the dynamics governed by~\eqref{eq:system}.
Hence, the resulting evolved solution $\bQ(\bx,t,\om)$ for $t > 0$ is also a random field, called \emph{random entropy solution}; see \cite{MSS11-syscl,Sukys14-dissertation,schwabsid1} for precise formulation and details.
%

\subsubsection{The classical MLMC}

The statistical moments of the QoIs, such as expectation $\IE[q]$, are obtained through sampling by the MLMC methodology.
Multi-level methods employ a hierarchy of spatial discretizations of the computational domain $D$, or, equivalently, a hierarchy of numerical deterministic solvers as described in \autoref{ss:fvm}, ordered by increasing ``level of accuracy'' $\ell = 0, \dots, L$.
On each such discretization level $\ell$, and for a given statistical realization (a ``sample'') $\om\in\Om$,
a numerical approximation of the QoI $q(\om)$ using the applied FVM will be denoted by $q_\ell(\om)$.

The classical  MLMC estimator \cite{GIL2} provides accurate and efficient estimates for statistical moments  of $q$ in terms of the telescoping sum of numerical approximations $q_\ell(\om)$ over all levels.
In particular, an approximation $\EMLMC [q_L]$ of $\IE[q]$ is constructed from the approximate telescoping sum
\begin{equation}
\label{eq:mlmc-sum}
\IE [q] \approx \IE [q_L] = \IE [q_0] + \sum_{\ell=1}^L \Big( \IE [q_\ell] - \IE [q_{\ell-1}] \Big)
\end{equation}
by additionally approximating all exact mathematical expectations using Monte Carlo sampling with \emph{level-dependent} number $M_\ell$ of samples for QoIs $q_\ell^i$,
leading to
\begin{equation}
\label{eq:mlmc}
\IE [q_L] \approx \EMLMC [q_L] =
\frac{1}{M_0} \sum_{i=1}^{M_0} q_0^i + \sum_{\ell=1}^L \frac{1}{M_\ell} \sum_{i=1}^{M_\ell} \Big( q_\ell^i - q_{\ell-1}^i \Big).
\end{equation}
We note that each sample $q^i_\ell$ is obtained by solving the governing system \eqref{eq:system} using the
FVM method from \autoref{ss:fvm} with discretization (number of mesh cells and time steps) corresponding to level $\ell$.
Assuming that samples for $q_0^i$ and differences $q_\ell^i - q_{\ell-1}^i$ are drawn independently (however, sample pairs $q_\ell^i$ and $q_{\ell-1}^i$ \emph{are} statistically dependent), the statistical mean square error of the (here assumed to be unbiased) standard MLMC estimator is given by the sum of sample-reduced variances of differences between every two consecutive levels,
\begin{equation}
\label{eq:mlmc-err}
\varepsilon^2 =
\IE \Big[ \Big( \EMLMC [q_L] - \IE[q_L] \Big)^2 \Big] =
\frac{\IV[q_0]}{M_0} + \sum_{\ell=1}^L \frac{\IV[q_\ell - q_{\ell-1}]}{M_\ell}.
\end{equation}
Assuming that $\IV [q_\ell] \approx \IV [q]$, the MLMC sampling error in \eqref{eq:mlmc-err} can be approximated in terms of correlation coefficients of every two consecutive levels, i.e.,
\begin{equation}
\label{eq:mlmc-err-corr}
\varepsilon^2 \approx \IV[q] \left( \frac{1}{M_0} + 2 \sum_{\ell=1}^L \frac{1 - \operatorname{Cor}[q_\ell,q_{\ell-1}]}{M_\ell} \right).
\end{equation}
Note, that strongly correlated QoIs on two consecutive levels lead to significant reduction in the required number of samples on levels $\ell > 0$.
Optimal number of samples $M_\ell$ for each level $\ell = 0, \dots L$ can be obtained
using empirical or approximate asymptotic estimates on $\IV[q_0]$ and $\IV[q_\ell - q_{\ell-1}]$
by minimizing the amount of total computational work $M_0\Work_0 + \dots + M_L \Work_L$ for a prescribed error tolerance $\tau$ such that $\varepsilon \leq \tau$ in \eqref{eq:mlmc-err}, as described in \cite{GIL2}.
Here, $\Work_{0}$ denotes the amount of computational work needed to compute a single sample (statistical realization) on a given resolution level $0$. For levels $\ell > 0$, $\Work_{\ell}$ denotes the amount of computational work needed to compute a pair of such samples on resolution levels $\ell$ and $\ell-1$.
Number of samples $M_\ell$ was shown to decrease exponentially with increasing level $\ell$,
and hence such reduction directly translates into large computational savings over single-level MC sampling methods,
as reported in \cite{schwabsid1,MSS11-syscl,MSS11-sw,SMS11-static,Sukys13-adaptive}.

\subsubsection{Optimal fidelity MLMC: control variates for two-level Monte Carlo}
\label{sss:cv-2lmc}

We present a novel method for reducing the variance and further increasing the efficiency of the classical MLMC method.
The backbone of MLMC is the hierarchical variance reduction technique.
Assuming only two levels, a coarser level $\ell - 1$ and a finer level $\ell$,
statistical moments at level $\ell$ use simulations from coarser discretization level $q_{\ell-1}$ as a \emph{control variate} with ``known'' $\IE[q_{\ell-1}]$ and the \emph{predetermined} coupling coefficient $\alpha_\ell = 1$. The coupled statistical QoI $q_\ell^*$ is given by:
\begin{equation}
\label{eq:cv}
q_\ell^* = q_\ell + \alpha \Big( \IE[q_{\ell-1}] - q_{\ell-1} \Big).
\end{equation}
The variance reduction that is achieved in \eqref{eq:cv} by replacing $q_\ell$ with $q^*_\ell$ depends on the correlation between $q_\ell$ and $q_{\ell-1}$,
\begin{equation}
\label{eq:cv-var}
\IV [ q_\ell^* ] =
\IV \Big[ q_\ell + \alpha \Big( \IE[q_{\ell-1}] - q_{\ell-1} \Big) \Big]
= \IV [q_\ell] + \alpha^2 \IV [q_{\ell-1}] -2 \alpha \Cov [q_\ell,q_{\ell-1}].
\end{equation}
The expectation $\IE[q_{\ell-1}]$ is considered ``known" since it can be approximated by sampling estimator $E_{M_{\ell-1}} [q_{\ell-1}]$ that is computationally affordable due to the lower resolution of the solver on coarser level $\ell-1$.
Statistical estimators using $M_\ell$ samples are applied to $q^*_\ell$ instead of $q_\ell$, leading to the building block of the MLMC estimator:
\begin{equation}
\label{eq:cv-est}
\IE [q_\ell] = \IE [q_\ell^*] \approx
E_{M_\ell,M_{\ell-1}} [q_\ell^*] =
E_{M_\ell} [q_\ell] + \alpha \Big( E_{M_{\ell-1}} [q_{\ell-1}] - E_{M_\ell} [q_{\ell-1}] \Big).
\end{equation}
In standard MLMC, the coefficient $\alpha$ in \eqref{eq:cv}--\eqref{eq:cv-est} is set to unity \cite{GIL2,Giles2015}. This constraint limits the efficiency of the variance reduction technique. In particular, assuming that variances at both levels are comparable, i.e. $\IV[q_\ell] \approx \IV[q_{\ell-1}]$, the standard MLMC estimator \emph{fails} to reduce the variance if the correlation coefficient $\operatorname{Cor} [q_\ell,q_{\ell-1}]$ drops below threshold of $1/2$,
\begin{equation}
\label{eq:cv-var-0.5}
\IV [ q_\ell^* ] \approx
2 \IV [q_\ell] - 2 \operatorname{Cor} [q_\ell,q_{\ell-1}] \IV [q_{\ell}] \geq
2 \IV [q_\ell] - 2 \frac{1}{2} \IV [q_\ell] = \IV[q_\ell].
\end{equation}
Hence, in standard MLMC, such moderate level correlations $\operatorname{Cor} [q_\ell,q_{\ell-1}] \leq 1/2$ would not provide any variance reduction; on the contrary, variance would be increased.
To avoid this, the optimal $\alpha$ minimizing the variance of $q_\ell^*$ as in \eqref{eq:cv-var} can be used:
\begin{equation}
\label{eq:alpha-opt}
\alpha = \frac{\Cov[q_\ell,q_{\ell-1}]}{\IV[q_{\ell-1}]} \approx \operatorname{Cor}[q_\ell,q_{\ell-1}].
\end{equation}
A consequence of \eqref{eq:alpha-opt} is that the \emph{predetermined} choice of $\alpha = 1$ in \eqref{eq:cv} is  optimal \emph{only} under very restrictive conditions: perfectly correlated levels with correlation coefficient $\operatorname{Cor}[q_\ell,q_{\ell-1}] = 1$ and assumption that coarser level estimates $\IE[q_{\ell-1}]$ are already available (hence no computation is needed, i.e.,~$\Work_{\ell-1} = 0$).
Note, that for optimal $\alpha$ as in \eqref{eq:alpha-opt}, variance is \emph{always}
reduced in \eqref{eq:cv-var}, even for $0 < \operatorname{Cor}[q_\ell,q_{\ell-1}] < 1/2$,
\begin{equation}
\label{eq:cv-var-opt}
\IV [ q_\ell^* ]
= \Big(1 - \operatorname{Cor} [q_\ell,q_{\ell-1}]^2 \Big) \IV [q_\ell] < \IV[q_\ell].
\end{equation}
For $\Work_{\ell-1} \neq 0$, it is necessary to obtain an 
estimate for $\IE[q_{\ell-1}]$ in \eqref{eq:cv} as well.
In such case, using the independence of estimators $E_{M_\ell}$ and $E_{M_{\ell-1}}$ and the central limit theorem, the variance of the two-level estimator $E_{M_\ell,M_{\ell-1}} [q_\ell^*]$ as in \eqref{eq:cv-est} is given by
\begin{equation}
\label{eq:cv-est-var}
\IV \big[ E_{M_\ell,M_{\ell-1}} [q_\ell^*] \big] =
\frac {\IV [q_\ell - \alpha q_{\ell-1}]} {M_\ell} +
\frac {\IV [\alpha q_{\ell-1}]} {M_{\ell-1}}.
\end{equation}
Given a constraint on the available computational budget $\cB$,
the number of available samples on each level depends on the required computation cost at that level, i.e.,
\begin{equation}
\label{eq:M-W}
M_\ell = \frac{\cB}{\Work_\ell}.
\end{equation}
Here, we have not yet considered the non-uniform distribution of computational budget $\cB$ across levels $\ell$ and $\ell-1$.
Such distribution will be addressed in terms of optimal number of samples $M_\ell$ for each level $\ell$ in the forth coming \autoref{sss:cv-mlmc}.
Plugging \eqref{eq:M-W} to \eqref{eq:cv-est-var} and multiplying by the computation budget $\cB$,
the total computational cost for variance reduction in $E_{M_\ell,M_{\ell-1}} [q_\ell^*]$ is given by
\begin{equation}
\label{eq:cv-2lmc-var}
C[{q^*_\ell}] =
\IV \big[ E_{M_\ell,M_{\ell-1}} [q_\ell^*] \big] \cB =
\IV [\alpha q_{\ell-1}] \Work_{\ell-1} + \IV [q_\ell - \alpha q_{\ell-1}] \Work_{\ell},
\end{equation}
where variances of $\alpha q_{\ell-1}$ and $q_\ell - \alpha q_{\ell-1}$ are weighted by the corresponding computational costs $\Work_{\ell-1}$ and $\Work_\ell$, respectively.
In order to find optimal $\alpha$,
the computational variance reduction cost $C[{q^*_\ell}]$ in \eqref{eq:cv-2lmc-var} is minimized instead of $\IV [ q_\ell^* ]$ in \eqref{eq:cv-var}.
The resulting optimal coefficient $\alpha$ is given by
\begin{equation}
\label{eq:alpha-opt-works}
\alpha = \frac{\Work_{\ell}}{\Work_{\ell} + \Work_{\ell-1}}\frac{\Cov[q_\ell,q_{\ell-1}]}{\IV[q_{\ell-1}]} \approx \frac{\Work_{\ell}}{\Work_{\ell} + \Work_{\ell-1}} \operatorname{Cor}[q_\ell,q_{\ell-1}].
\end{equation}
%
We note, that \eqref{eq:alpha-opt-works} reduces to standard control variate coefficient \eqref{eq:alpha-opt} for $\Work_{\ell-1} = 0$.

\subsubsection{Optimal fidelity MLMC: control variates for multi-level Monte Carlo}
\label{sss:cv-mlmc}

We generalize the concept of optimal control variate coefficients from \autoref{sss:cv-2lmc}
to arbitrary number of levels in MLMC estimator.
In particular, control variate contributions from multiple levels can be used,
\begin{equation}
\label{eq:cv-mlmc}
q_L^* =
\Big( q_L - \alpha_{L-1} q_{L-1} \Big) + \Big( \alpha_{L-1} q_{L-1} - \alpha_{L-2} q_{L-2} \Big) + \dots + \Big( \alpha_1 q_1 - \alpha_0 q_0 \Big) + \alpha_0 q_0,
\end{equation}
allowing for generalization of the telescoping sum in \eqref{eq:mlmc-sum} to the control variate setting introduced in \eqref{eq:cv}, where the coefficient for the finest level $\ell = L$ is fixed to $\alpha_L = 1$:
\begin{equation}
\label{eq:cv-mlmc-sum}
\IE [q_L] = 
\IE [q_L^*] = 
\alpha_0 \IE [q_0] + \sum_{\ell=1}^L \Big( \alpha_\ell \IE [q_\ell] - \alpha_{\ell-1} \IE [q_{\ell-1}] \Big).
\end{equation}
Using the mutual statistical independence of $\alpha_0 q_0$ and differences $\alpha_\ell q_\ell - \alpha_{\ell-1} q_{\ell-1}$ together with the central limit theorem analogously as in \autoref{sss:cv-2lmc}, the total computational variance reduction cost, generalizing \eqref{eq:cv-2lmc-var} for $q_L^*$ over all levels as in \eqref{eq:cv-mlmc}, is given by
\begin{equation}
\label{eq:cv-mlmc-var}
\begin{aligned}
C [ q_L^* ]
&=\IV[\alpha_0 q_0] \Work_0 + \sum_{\ell=1}^L \IV[\alpha_\ell q_\ell - \alpha_{\ell-1} q_{\ell-1}] \Work_\ell \\
&= \alpha_0^2\IV [q_0] \Work_0 + \sum_{\ell=1}^L \Big( \alpha_\ell^2 \IV [q_\ell] + \alpha_{\ell-1}^2 \IV [q_{\ell-1}] - 2 \alpha_\ell \alpha_{\ell-1} \Cov [q_\ell,q_{\ell-1}] \Big) \Work_\ell.
\end{aligned}
\end{equation}
Minimization of $C [ q_L^* ]$ pertains to solving a linear system of equations,
\begin{equation}
\label{eq:alpha-eqsys}
\frac{\partial}{\partial\alpha_\ell} C [ q_L^* ] = 0, \qquad \ell = 0, \dots, L-1.
\end{equation}
Denoting $\sigma^2_\ell = \IV [q_\ell]$, $\Work_{\ell,\ell-1} = \Work_{\ell} + \Work_{\ell-1}$, and $\sigma^2_{\ell,\ell-1} = \Cov [q_\ell,q_{\ell-1}]$, system \eqref{eq:alpha-eqsys} can be written
in a form of a tridiagonal matrix
\begin{equation}
\label{eq:alpha-matvec}
\begin{bmatrix}
       \sigma_0^2 \Work_{1,0} \hspace*{-12pt} & -\sigma^2_{1,0} \Work_1 \hspace*{-12pt} & ~ \hspace*{-12pt} & ~ \hspace*{-12pt} \\
       -\sigma^2_{1,0} \Work_1 & \ddots & \ddots & ~ \\
       ~ & \ddots & \ddots & -\sigma^2_{L-1,L-2} \Work_{L-1} \\
       ~ & ~ &-\sigma^2_{L-1,L-2} \Work_{L-1} & \sigma_{L-1}^2 \Work_{L,L-1}
\end{bmatrix}
\begin{bmatrix}
       \alpha_0 \\
       \alpha_1 \\
       \vdots \\
       \alpha_{L-2} \\
       \alpha_{L-1}
\end{bmatrix}
=
\begin{bmatrix}
       0 \\
       0 \\
       \vdots \\
       0 \\
       \sigma^2_{L,L-1} \Work_L
\end{bmatrix}\hspace*{-3pt}.
\end{equation}
For a special case of two levels with $L=1$, the
solution of \eqref{eq:alpha-matvec} reduces to \eqref{eq:alpha-opt-works}.
Generalizing \eqref{eq:mlmc}, the unbiased \emph{Optimal Fidelity Multi-Level Monte Carlo} (OF-MLMC) estimator is then given by
\begin{equation}
\label{eq:ocv-mlmc}
\IE [q_L] \approx \EOFMLMC [q_L] =
\alpha_0 \frac{1}{M_0} \sum_{i=1}^{M_0} q_0^i + \sum_{\ell=1}^L \frac{1}{M_\ell} \sum_{i=1}^{M_\ell} \Big( \alpha_\ell q_\ell^i - \alpha_{\ell-1} q_{\ell-1}^i \Big),
\end{equation}
with optimal control variate coefficients $\alpha_\ell$ given by \eqref{eq:alpha-matvec}.
We assume  that samples for $\alpha_0 q_0^i$ and differences $\alpha_\ell q_\ell^i - \alpha_{\ell-1} q_{\ell-1}^i$ are drawn independently (however, individual $q_\ell^i$ and $q_{\ell-1}^i$ \emph{are} statistically dependent). The  statistical mean square error $\varepsilon^2_{\text{OF}}$ of the OF-MLMC estimator is given by the sum of sample-reduced variances of $\alpha_\ell$-weighted differences between every two consecutive levels,
\begin{equation}
\label{eq:ocv-mlmc-err}
\begin{aligned}
\varepsilon^2_{\text{OF}} =
\IE \Big[ \Big( \EOFMLMC [q_L] - \IE[q_L] \Big)^2 \Big]
&= \frac{\IV[\alpha_0 q_0]}{M_0} + \sum_{\ell=1}^L \frac{\IV[\alpha_\ell q_\ell - \alpha_{\ell-1} q_{\ell-1}]}{M_\ell} \\
&=: \hspace{5pt} \frac{\tilde\sigma^2_0}{M_0} \quad + \quad \sum_{\ell=1}^L \frac{\tilde\sigma^2_\ell}{M_\ell}.
\end{aligned}
\end{equation}
Given computational costs $\Work_\ell$ of evaluating a single approximation $q_\ell^i$ for each level $\ell = 0, \dots, L$ and
a desired tolerance $\tau > 0$, the total computational cost of OF-MLMC can be minimized under constraint of
$\varepsilon^2_{\text{OF}} \leq \tau$ by choosing optimal
number of MC samples $M_\ell$ on each level according to \cite{Florian}, yielding
\begin{equation}
\label{eq:M-tol}
M_\ell = \left\lceil \frac{1}{\tau^2} \sqrt{\frac{\tilde\sigma_\ell^2}{\Work_\ell}} \sum_{k=0}^L \sqrt{\tilde\sigma_k^2\Work_k} \right\rceil.
\end{equation}
Alternatively, given available total computational budget $\mathcal{B}$ instead of a desired tolerance $\tau$,
the OF-MLMC error $\varepsilon^2_{\text{OF}}$ is minimized by choosing optimal $M_\ell$ according to
\begin{equation}
\label{eq:M-budget}
M_\ell = \left\lceil \mathcal{B} \sqrt{\frac{\tilde\sigma_\ell^2}{\Work_\ell}} \Big/ \sum_{k=0}^L \sqrt{\tilde\sigma_k^2\Work_k} \right\rceil.
\end{equation}

\subsubsection{Adaptive optimal fidelity MLMC algorithm}
\label{sss:ocv-mlmc-adaptive}

The OF-MLMC  algorithm proceeds iteratively, with each iteration improving the accuracy of the estimated statistics for the quantities of interest, such as the expectation $\IE[q]$. Each iteration also improves the accuracy of auxiliary parameters, such as $\sigma_\ell^2$, $\sigma^2_{\ell,\ell-1}$, $\Work_\ell$, $\tilde\sigma_\ell^2$, and the optimality of the number of samples $M_\ell$ for each level $\ell = 0, \dots, L$.\\
A single iteration of the algorithm consists of the following 8 steps.

\begin{enumerate}
\item {\bfseries Warm-up:} Begin with \emph{level-dependent} number of warm-up samples,
\begin{equation}
\label{eq:warmup}
M_\ell = \left\lceil \frac{\Work_L}{\Work_\ell} \frac{1}{2^{(L-\ell)}} \right\rceil, \quad \ell = L, L-1, \dots, 0.
\end{equation}
The choice of $M_\ell$ as in \eqref{eq:warmup}
prevents efficiency deterioration of the final OF-MLCM estimator
by ensuring that the total computational budget for the warm-up iteration does not exceed $2\Work_L$;
at the same time, it allows to prescribe sufficiently many samples on the coarser levels, where $\tilde\sigma_\ell^2$ is expected to be large.
In our particular application, computational work for each level scales as $\Work_\ell = \cO(2^{4\ell})$,
and hence the amount of required warm-up samples is given by
$ 1, 8, \dots, 2^{3L}$ for $  \ell = L, L-1, \dots, 0$.
We note, that constant (level-independent) number of warm-up samples can be very inefficient \cite{Florian,Tempone}.
%
%
\item {\bfseries Solver:} Evaluate approximations $q_\ell^i$ for each level $\ell = 0, \dots, L$ and sample $i = 1, \dots, M_\ell$.
\item {\bfseries Indicators:} Using $q_\ell^i$, estimate $\sigma_\ell^2$, $\sigma^2_{\ell,\ell-1}$, and $\Work_\ell$ for $\ell = 0, \dots, L$.
Optionally, empirical estimates of $\sigma^2_{\ell,\ell-1}$
could be used within Bayesian inference framework to fit an exponential decay model for $\sigma^2_{\ell,\ell-1}$ w.r.t. levels $\ell = 1, \dots, L-1$.
Assuming Gaussian errors, this reduces to a least-squares line fit to the natural logarithm of indicators $\sigma^2_{\ell,\ell-1}$.
\item {\bfseries Coefficients:} Compute control variate coefficients $\alpha_\ell$ from estimated $\sigma_\ell^2$ and $\sigma^2_{\ell,\ell-1}$ using \eqref{eq:alpha-matvec}.
\item {\bfseries Errors:} Using $q_\ell^i$ and $\alpha_\ell$, estimate $\alpha_\ell$-weighted covariances $\tilde\sigma_\ell^2$
and total sampling error $\hat\varepsilon^2_{\text{OF}} \approx \varepsilon^2_{\text{OF}}$ as in \eqref{eq:ocv-mlmc-err}.
\item {\bfseries Estimator:} If the required tolerance is reached,
i.e.,~$\hat\varepsilon_{\text{OF}} \leq \tau$,
or if the prescribed computational budget $\mathcal{B}$ is spent,
then evaluate OF-MLMC estimator \eqref{eq:ocv-mlmc} and terminate the algorithm.
Otherwise, proceed to the optimization step.
\item {\bfseries Optimization:} Compute optimal required number of samples $\hat M_\ell$ from $\tilde\sigma_\ell$ and $\Work_\ell$
using either \eqref{eq:M-tol} or \eqref{eq:M-budget}, respectively.\\
%
\emph{Remark.} If we obtain $\hat M_\ell < M_\ell$  for some level $\ell$,
we \emph{keep} the already computed samples,
i.e.,~we set $\hat M_\ell = M_\ell$.
In order to adjust for such a constraint in the optimization problem,
we subtract the corresponding sampling error $\tilde\sigma^2_\ell/M_\ell$ from the
required tolerance $\tau^2$, or subtract the corresponding amount of computational budget $M_\ell\Work_\ell$ from $\mathcal{B}$, respectively.
Afterwards, the number of samples $\hat M_\ell$ for the \emph{remaining} levels (where $\hat M_\ell = M_\ell$ was not enforced) are re-optimized according to either \eqref{eq:M-tol} or \eqref{eq:M-budget}, respectively.
We repeat this procedure until $\hat M_\ell \geq M_\ell$ is satisfied for all levels $\ell = 0, \dots, L$.
\item {\bfseries Iterations:} Go back to step (2) and continue the algorithm with the updated number of samples $\hat M_\ell$.
\end{enumerate}
If the empirical estimates in steps (3)--(5) of the above adaptive OF-MLMC algorithm are sufficiently accurate,
it will terminate after two iterations - the initial warm-up iteration and one additional iteration with already optimal $\alpha_\ell$ and $M_\ell$.
%
%
A more detailed discussion of the special cases within OF-MLMC algorithm,
alternative approaches, related work and possible extensions is provided in \autoref{a:discussion}.

\subsection{PyMLMC}

\begin{figure}[ht]
	\centering
	\includegraphics[width=0.8\textwidth]{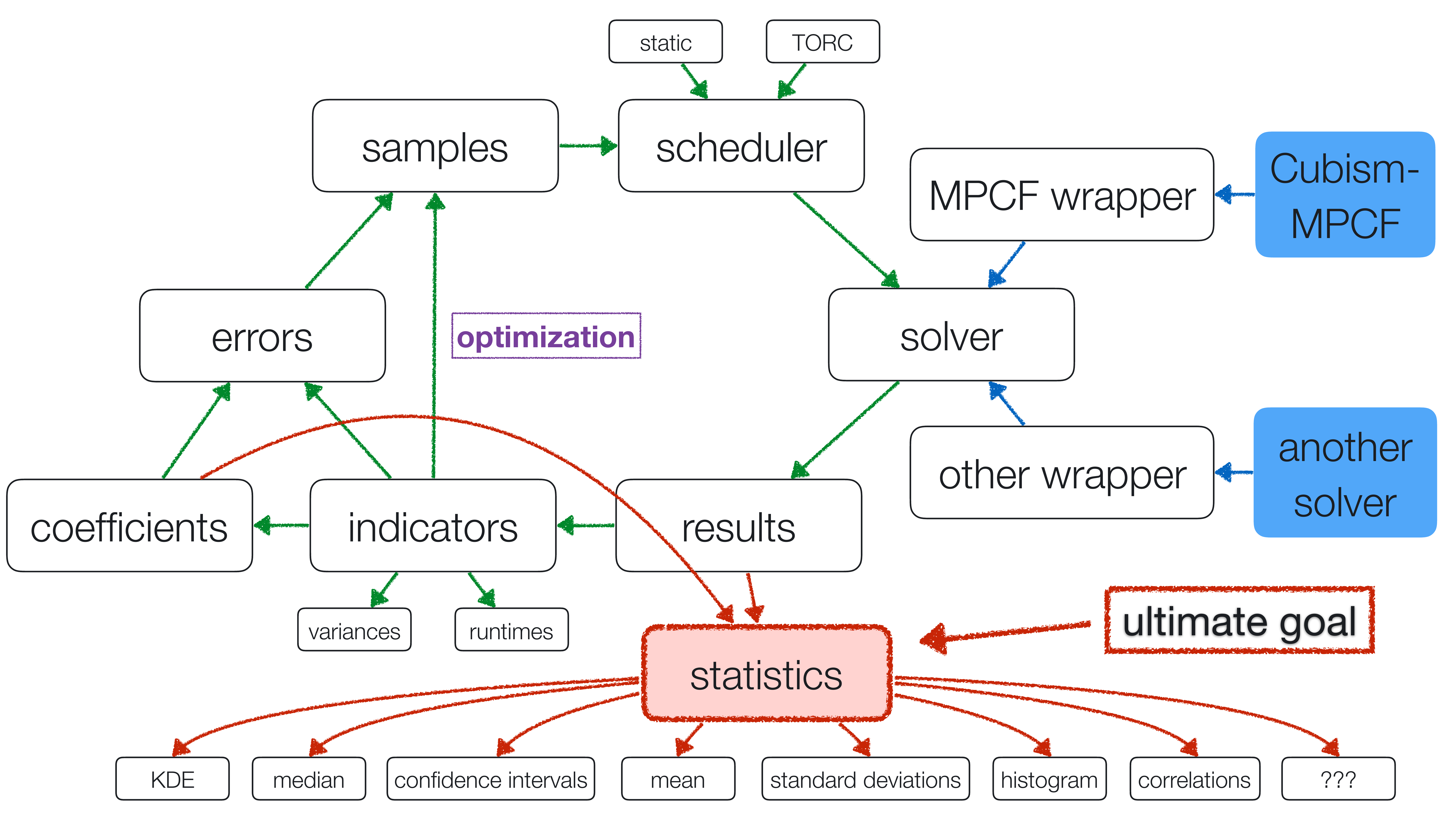}
	\caption[Scheme of the OF-MLMC implementation library PyMLMC]
	{Scheme of the OF-MLMC implementation library PyMLMC.}
	\label{f:pymlmc}
\end{figure}

The OF-MLMC algorithm is distributed as an open source library PyMLMC \cite{pymlmc}. A diagram of the software components is shown in \autoref{f:pymlmc}.
PyMLMC provides a modular framework for native execution of  deterministic solvers in their respective ``sandbox'' directories. This allows 
 maximum flexibility for programming languages, distributed and/or shared memory architectures, and support for many-core accelerators.
Due to the lack of communication among such sandboxed executions for each sample, the load balancing across samples can be relayed to the submission system (e.g. Slurm, LSF, LoadLeveler, Cobalt) of the compute cluster. Nested (across and within samples) parallelism is used, where few large parallel jobs are submitted for fine levels, and, in parallel, many small (possibly even serial) jobs are submitted for coarse levels.
To increase the efficiency and reduce the stress on submission systems, job batching (submitting a single job that computes multiple samples subsequently) and job merging (submitting a single job that computes multiple samples concurrently) or a combination of both is implemented.
Once all samples (or at least some of them) are computed, statistical estimators are constructed as a post-processing step using the NumPy and SciPy libraries. The ``sandboxing'' framework enables any additional required statistical estimates or QoIs to be evaluated at a later stage without the need to re-execute any of the computationally expensive sampling runs.
The amount of required disk space in multi-level methods scales linearly w.r.t. the amount of required computational budget. In particular, the total required disk space for all samples on all levels is of the same order as a \emph{single} statistical estimator on the full three-dimensional domain.
Hence, it is highly advantageous to keep all computed samples for the aforementioned post-processing flexibility purposes.

In the present work (see \autoref{s:results}), we verified the efficiency of the  (nested) parallelization of the OF-MLMC coupled with the Cubism MPCF solver on the entire MIRA supercomputer (Argonne National Laboratory) consisting of half a million cores. We note that large (exa-)scale simulations on massively parallel computing platforms are subject to  processor failures at run-time \cite{Cap09}. Exploiting the natural fault tolerance in OF-MLMC-FVM due to independent sampling, a Fault Tolerant MLMC (FT-MLMC) method was implemented in \cite{Stefan2} and was shown to perform in
agreement with theoretical analysis in the presence of simulated, compound Poisson distributed, random hard failures of
compute cores. Such FT-mechanisms are also available in PyMLMC, and have successfully excluded one corrupted sample on the coarsest level in the simulations reported in \autoref{s:results}.

\section{Numerical simulations and analysis}
\label{s:results}

The initialization of the cavities employs experimental findings indicating log-normal distribution for their  radii \cite{Morch:1982},
whereas the position vectors are generated according to a uniform distribution as there is no prior knowledge.

\subsection{Spherical cloud of 500 gas cavities with log-normally distributed radii}

\begin{figure}[ht]
	\centering
	\includegraphics[width=0.49\textwidth]{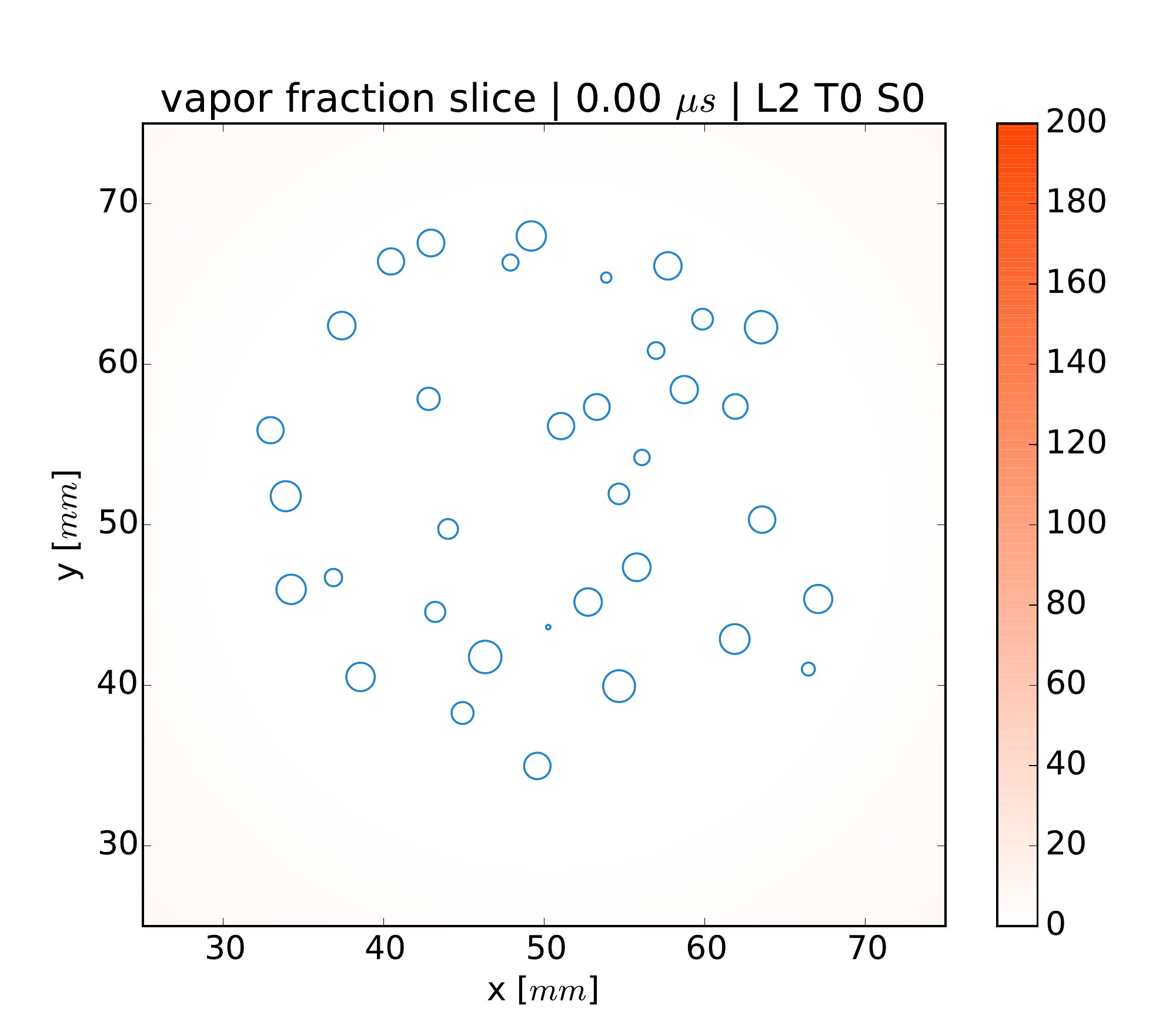}
	\includegraphics[width=0.49\textwidth]{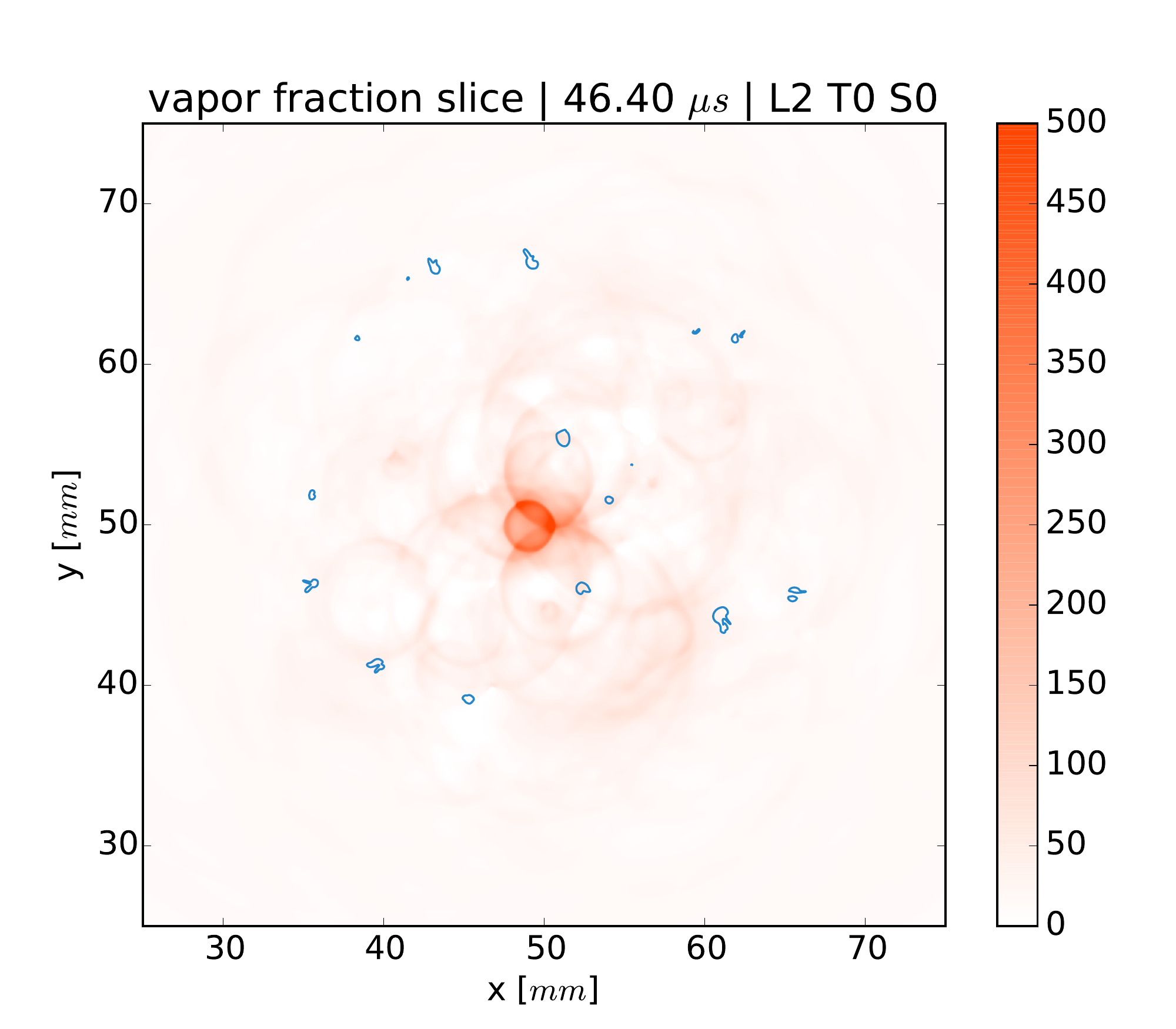}
	\caption[Sample of the pressure slice with cavity interfaces]
	{Single statistical realization (sample) of the pressure field and cavity interfaces in the two-dimensional slice at $z=50\ mm$ at the initial (left, $t = 0\ \mu s$) and at the collapse (right, $t = 46.4\ \mu s$) stages. Cavities collapse inwards, emitting pressure waves that focus near the center of the cloud. There is a two orders of magnitude difference in pressures at initial and collapse stages.}
	\label{f:uqvf500lFF_sample_slice_a}
\end{figure}

For a cubic domain $D = [0\ mm,100\ mm]^3$, we consider a cloud of $500$ bubbles located at the center $(50\ mm, 50\ mm, 50\ mm)^\top$ of the domain with a radius of $R_\text{cloud} = 20\ mm$.
The log-normal distribution for the radii of the cavity is clipped so as to contain bubbles only within the range of $r_{\min} = 0.8\ mm$ to $r_{\max} = 1.2\ mm$.
%
The resulting cloud gas volume content (w.r.t. to the volume of the sphere with radius $R_\text{cloud}$) is approximately $5\%$ and the resulting cloud interaction parameter $\beta$ is approximately $3$, where $\beta = \alpha (\frac{R_\text{cloud}}{R_\text{avg}})^2$ with cloud gas volume fraction $\alpha$, cloud radius $R_\text{cloud}$ and average cavity radius $R_\text{avg}$ (refer to \cite{Brennen:1997} for a derivation). We note that both of these quantities depend on a statistical realization of the random cloud.
An illustration of the cloud geometry is shown in \autoref{f:uqvf500lFF_sample_slice_a}.

The cloud is initially at pressure equilibrium with the surrounding water phase
at $p_{2,0} = 0.5\ MPa$. Throughout the entire domain, the density of the gas phase is set to $\rho_{2,0} = 5\ kg/m^3$ and the density of the liquid is set to $\rho_{1,0} = 1000\ kg/m^3$.
Starting $1\ mm$ away from the surface of the cloud, there is a smooth pressure increase towards the prescribed ambient pressure of $p_\infty = 10\ MPa$,
following the setup proposed in \cite{Tiwari:2013}.
The resulting pressure ratio is $p_\infty / p_{2,0} = 20$.
At the boundaries, non-reflecting, characteristic-based boundary conditions are applied, together with a penalty term
for the prescribed far-field pressure of $p_\infty = 10\ MPa$  \cite{Poinsot:1992}.
A single statistical realization (sample) of the pressure field and cavity interfaces across the two-dimensional slice at $z=50\ mm$ computed using Cubism-MPCF with resolution of $1024^3$ mesh cells is depicted in \autoref{f:uqvf500lFF_sample_slice_a}.
Cavities are observed to collapse inwards, emitting pressure waves that focus near the center of the cloud at time $t = 46.4\ \mu s$.

We consider four levels of spatial and temporal resolutions. The 
coarsest mesh consists of $512^3$ cells with two intermediate meshes of $1024^3$ and $2048^3$ resolutions, and 
the finest mesh with $4096^3$ cells. The time-step size decreases according to a prescribed CFL condition with CFL number set to $0.3$, resulting in approximately $2'250$ and $20'000$ time steps for the coarsest and finest levels, respectively.

We note that the number of uncertainty sources
in this simulation is very large: for each realization of a cloud, random three-dimensional spatial coordinates together with a random positive radius for all $500$ cavities are needed, comprising in total $2'000$ independent sources of uncertainty.

For each statistical sample of a collapsing cloud configuration and on each resolution level,
simulations were performed for approximately $70\ \mu s$ in physical time.
Depending on the random configuration of the cloud,
the main collapse occurred at approximately $40 - 50\ \mu s$,
followed by rebound and relaxation stages after $50\ \mu s$.
The obtained results are discussed in the following sections.

\subsection{Performance of the OF-MLMC}

We  quantify the computational gains  of the OF-MLMC method
by comparing it to standard MLMC and plain MC sampling methods.
The chosen quantity of interest $q$ for \eqref{eq:cv-mlmc-sum} is the pressure as sampled by a sensor $p_\bc$ placed at the center of the cloud and emulated as 
\begin{equation}
\label{eq:sensor}
p_\bc (t) = \frac{1}{|\cB_r(\bc)|}\int_{\cB_r(\bc)} p(\bx,t) \ d \bx,
\end{equation}
where pressure $p$ is averaged over a sphere $\cB_r(\bc)$ around the center of the cloud located at $\bc = (50\ mm, 50\ mm, 50\ mm)^\top$ with radius $r = 0.5\ mm$,
\begin{equation}
\label{eq:sensor-params}
\cB_r (\bc) = \{ \bx \in D : \| \bx - \bc \|_2 \leq r \}.
\end{equation}

\begin{figure}[ht]
  \centering
  \includegraphics[width=0.49\textwidth]{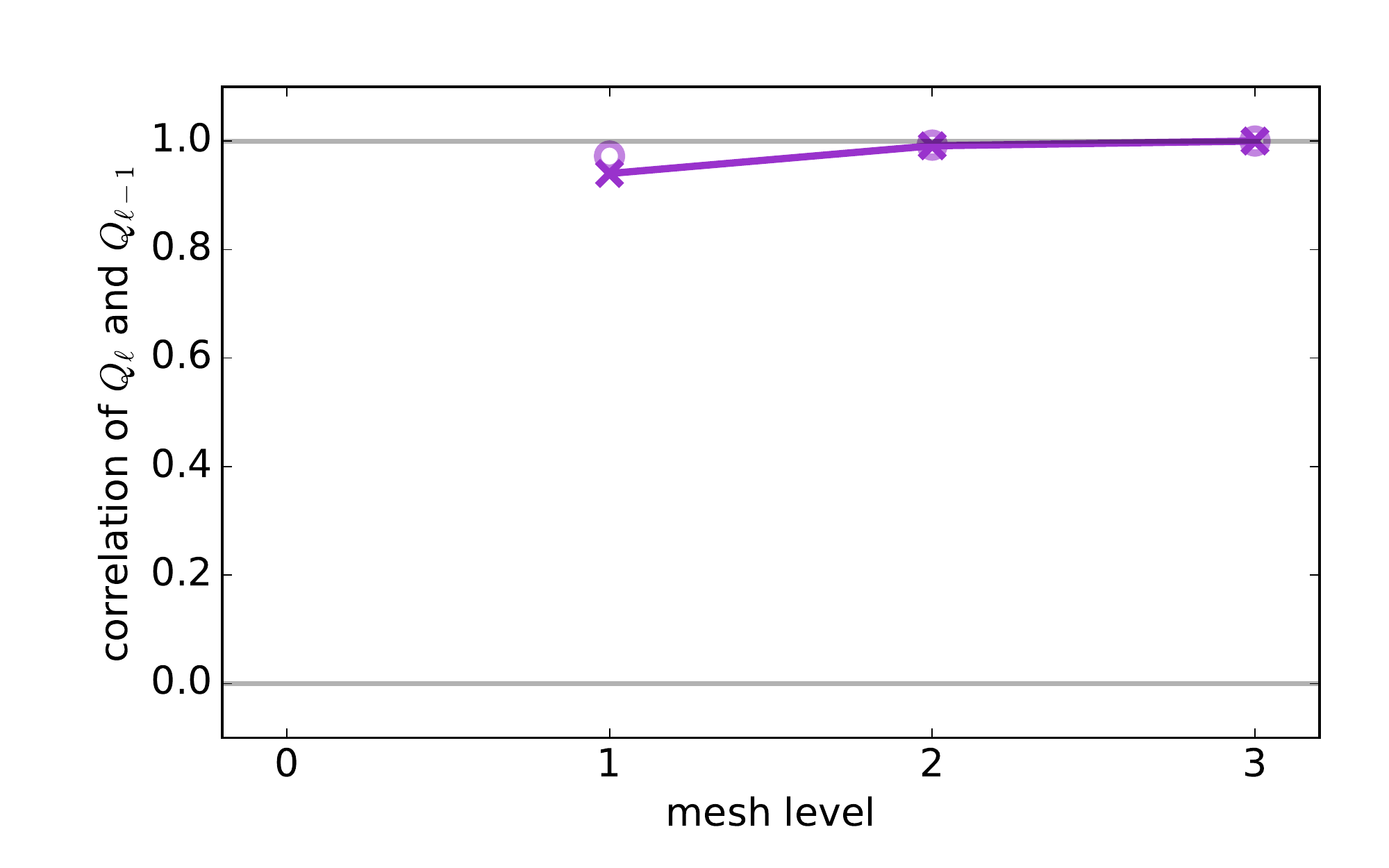}
  \includegraphics[width=0.49\textwidth]{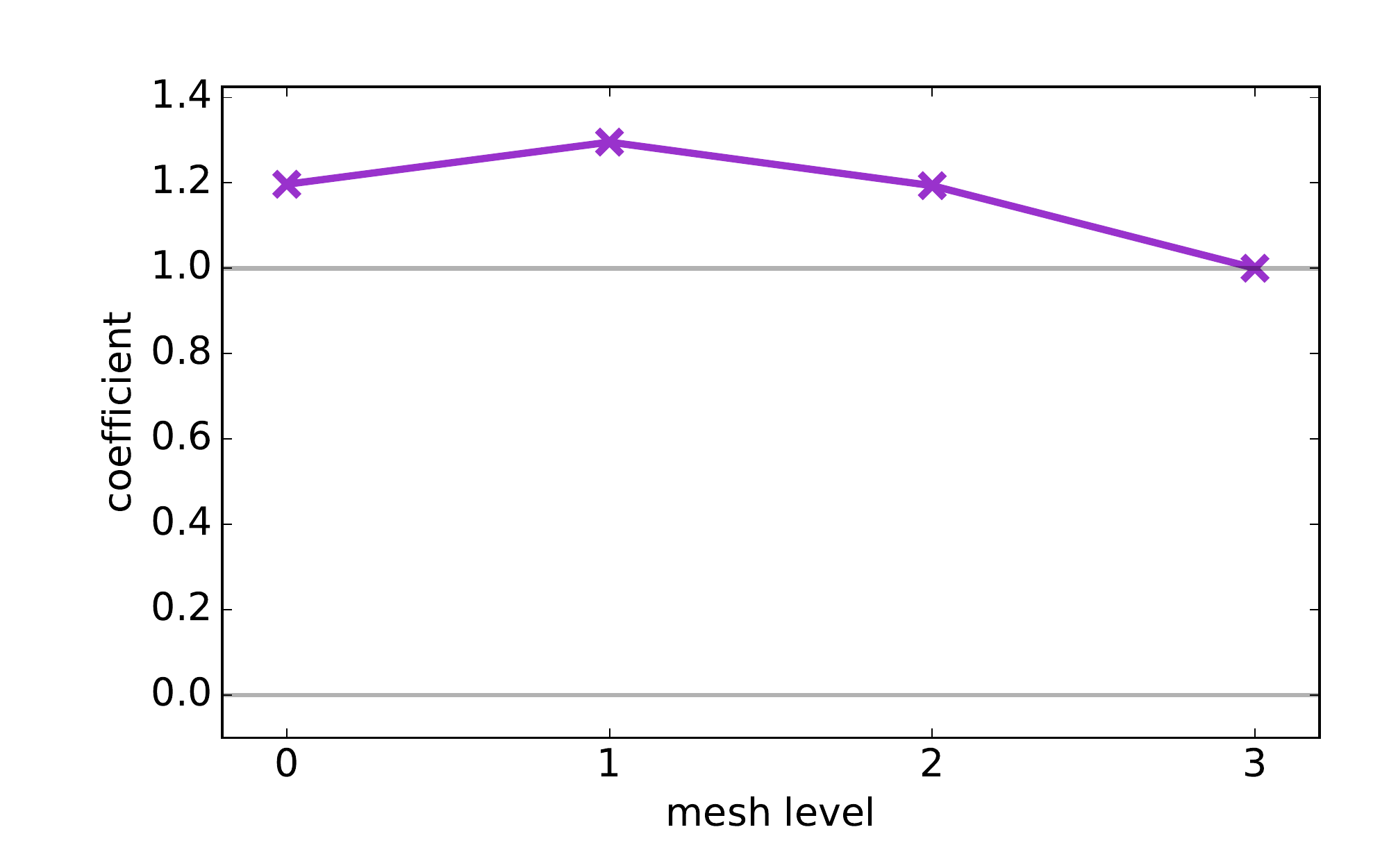}
  \caption[Estimated correlations and optimal control variate coefficients]
            {Estimated correlations between levels (left) and the resulting optimal control variate coefficients (right) as in \eqref{eq:alpha-matvec}.
            In left plot, circles depict measurements of correlations
            with the associated error bars, and solid line depicts the resulting inferred values.
            Relatively high correlations and moderate deviations (up to 30\%) in optimal control variate coefficients are observed.}
  \label{f:uqvf500lFF_correlations}
\end{figure}

For this particular choice of the QoI $q = p_\bc$,
estimated correlations between levels, implicitly used in \eqref{eq:alpha-matvec}, and the resulting optimal control variate coefficients from \eqref{eq:alpha-matvec}
are depicted in \autoref{f:uqvf500lFF_correlations}.
Due to relatively high correlations between resolution levels,
optimal control variate coefficients exhibit only moderate deviations from unity,
with the largest being at level 1 with deviation of 30\%.

\begin{figure}[ht]
  \centering
  \includegraphics[width=0.49\textwidth, trim = 570 0 0 0, clip]{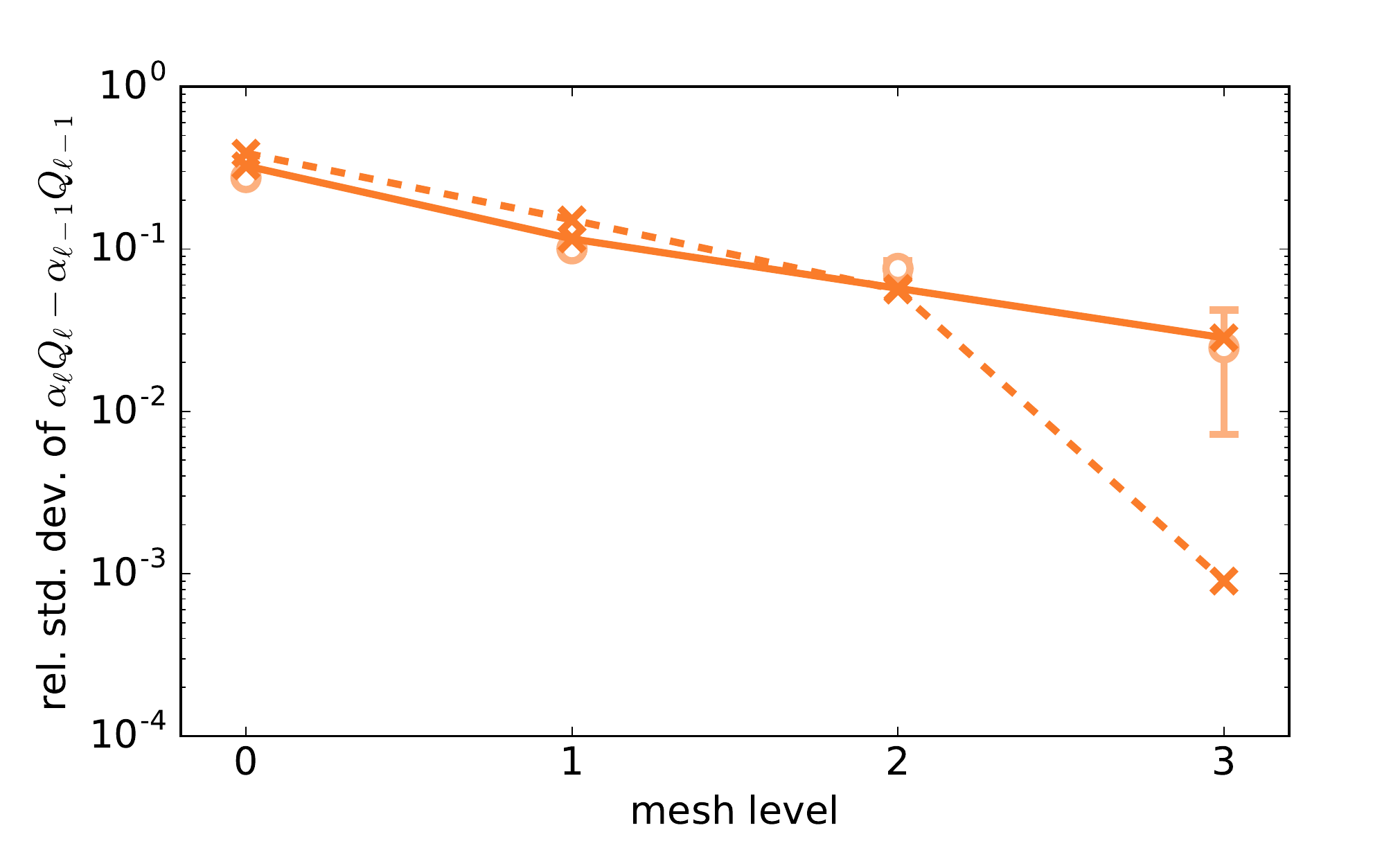}
  \includegraphics[width=0.49\textwidth]{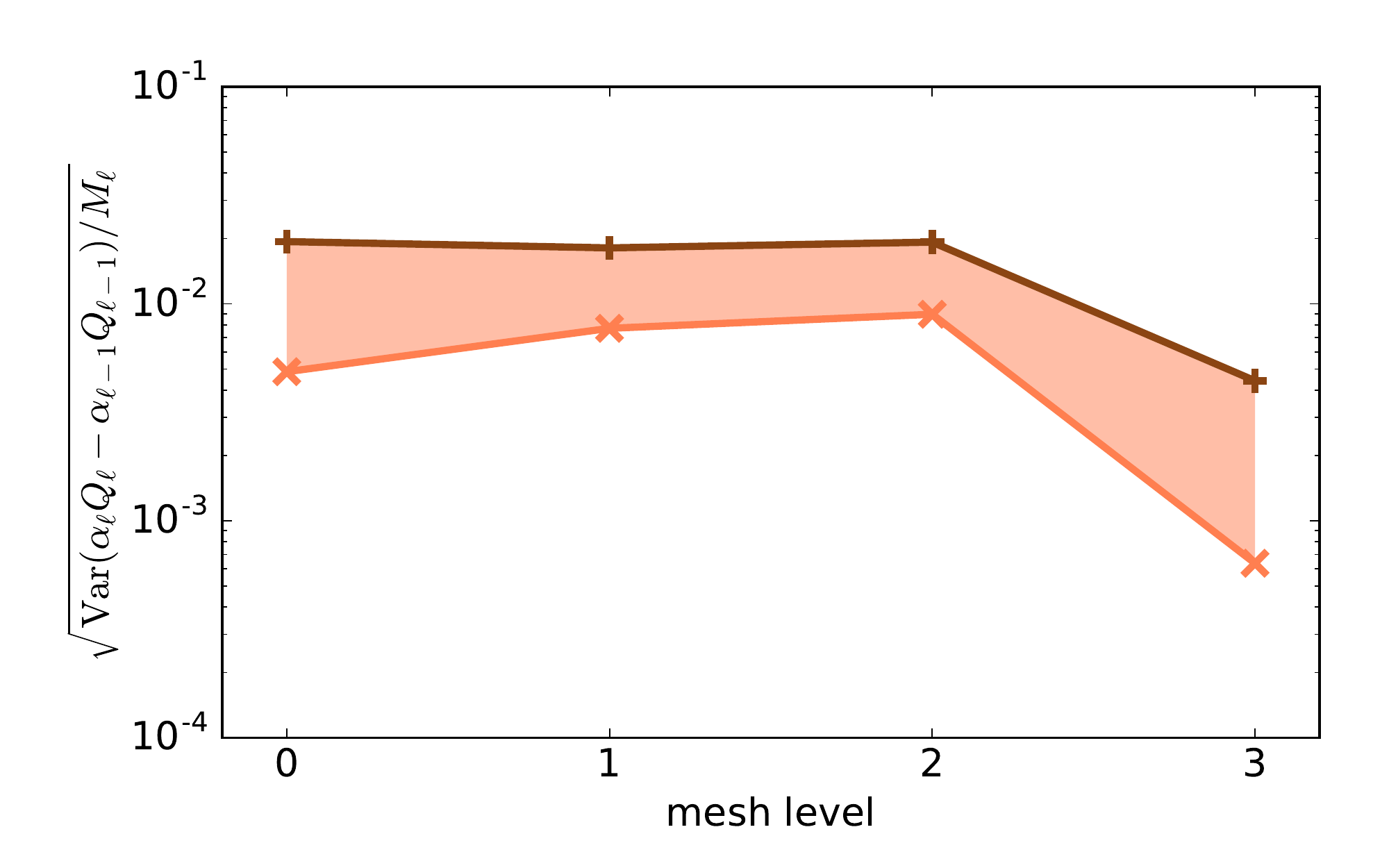}
  \caption[Estimated variances of differences and and resulting sampling errors]
            {Estimated variances of differences $\sigma^2_{\ell,\ell-1}$ (left) required in \eqref{eq:alpha-matvec}
            and the resulting estimated sampling errors $\varepsilon_\ell$ (right) defined in \eqref{eq:ocv-mlmc-err} on each level.
            In left plot, circles depict measurements $\sigma^2_{\ell,\ell-1}$
            with the associated error bars, solid line depicts the resulting inferred values,
            and dashed line depicts the adjusted values $\tilde \sigma^2_{\ell,\ell-1}$
            from \eqref{eq:ocv-mlmc-err}, where optimal control variate coefficients are applied.
            In right plot, brown line indicates estimated errors on each level after the initial MLMC warmup iteration,
            whereas orange line and the corresponding region indicate the estimated error improvements after the final MLMC iteration.
            The final total statistical error is estimated to be approximately $1.3 \cdot 10^{-2}$.}
  \label{f:uqvf500lFF_indicators_errors}
\end{figure}

Estimated variances of level differences, required in \eqref{eq:alpha-matvec},
and sampling errors for each level, computed in \eqref{eq:ocv-mlmc-err},
are depicted in \autoref{f:uqvf500lFF_indicators_errors}.
Variances of differences are decreasing for finer levels of resolution, which requires a smaller number of MC samples $M_\ell$ in order to reduce the statistical error on finer resolution levels, where sampling entails computationally very expensive numerical simulations.
Measurements of variances of differences $\sigma^2_{\ell,\ell-1}$
are plotted as circles, with the associated error bars, estimated from the variance of the estimator and the number of samples used in the warm-up stage.
These measurements, together with assumed Gaussian error model from the error bars, are used within Bayesian inference framework to fit an exponential decay model for $\sigma^2_{\ell,\ell-1}$ w.r.t. ``difference'' levels $\ell = 1, 2, 3$.
Solid line depicts the resulting inferred values, which are later used in \eqref{eq:alpha-matvec},
whereas dashed line depicts the adjusted values $\tilde \sigma^2_{\ell,\ell-1}$
from \eqref{eq:ocv-mlmc-err}, where optimal control variate coefficients are applied.
The resulting statistical errors at both MLMC iterations (warm-up and final) are decreasing w.r.t. the increasing resolution level. At the final MLMC iteration, the errors are significantly decreased on all levels when compared to the warm-up iteration, resulting in the total statistical error estimate of approximately $1.3 \cdot 10^{-2}$.

\begin{figure}[ht]
  \centering
  \includegraphics[width=0.49\textwidth]{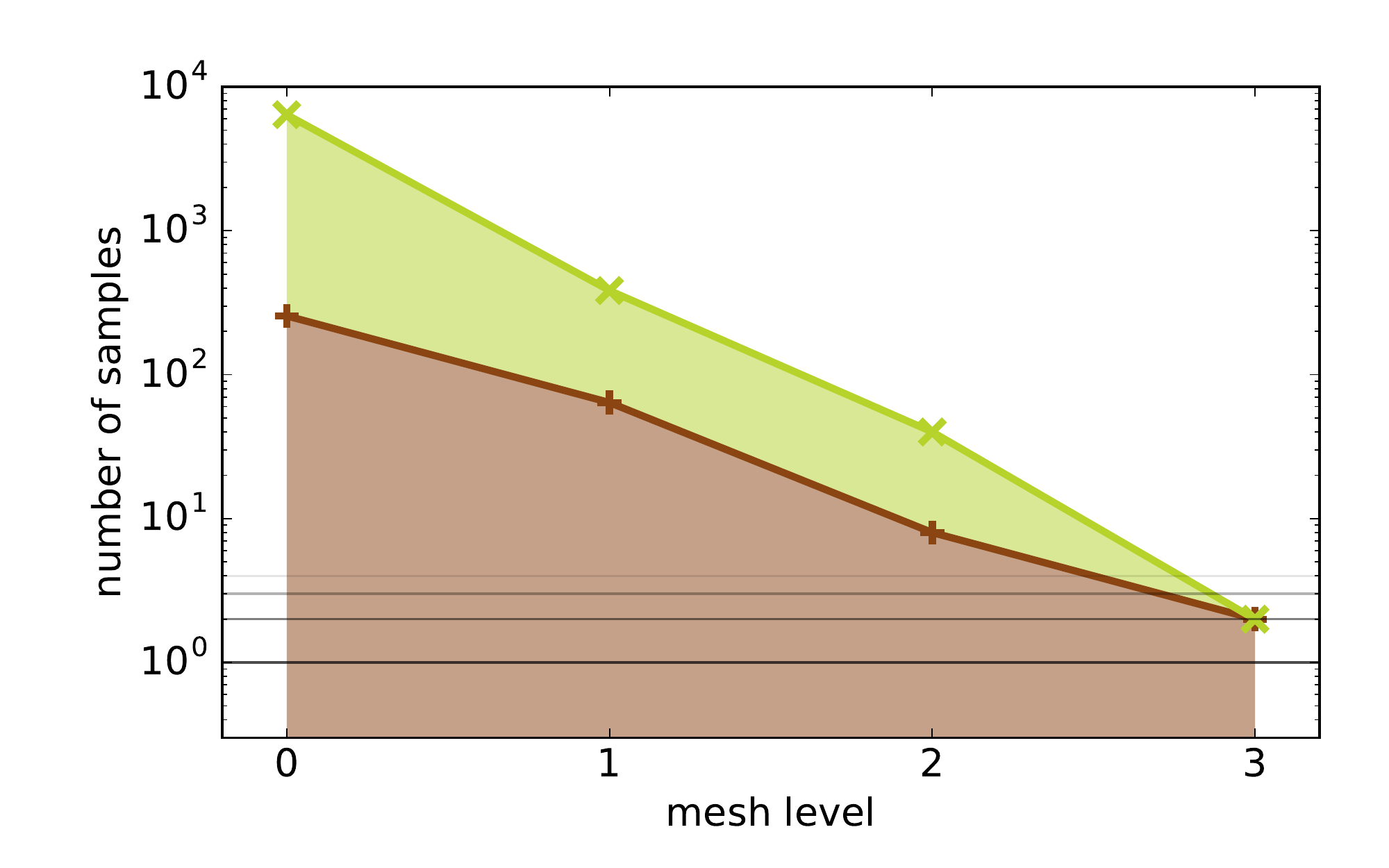}
  \includegraphics[width=0.49\textwidth]{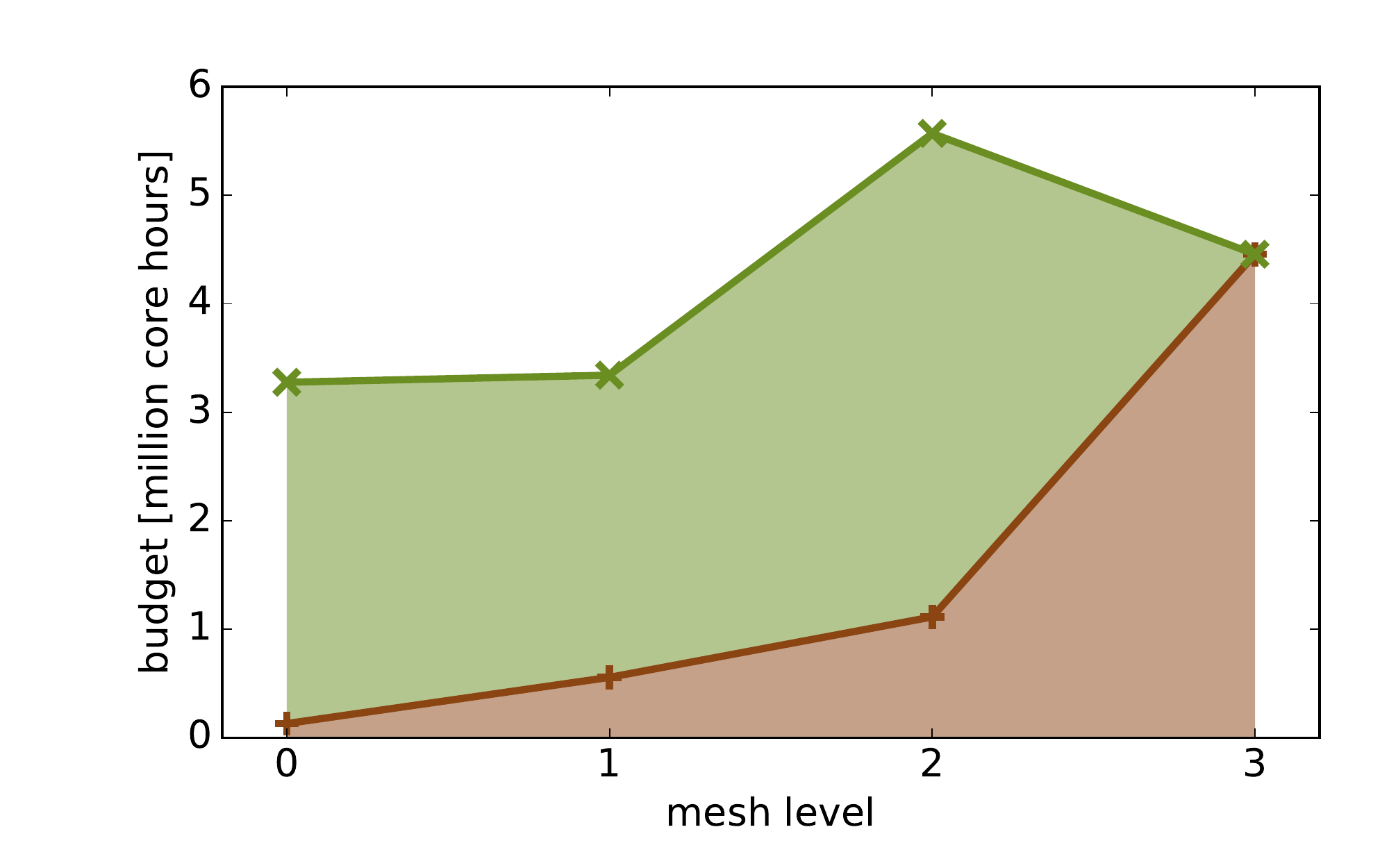}
  \caption[Estimated number of samples and resulting sampling errors]
            {Estimated number of samples $M_\ell$  for the warm-up and final iterations of the OF-MLMC algorithm (left) and the resulting estimated computational budget (right) on each level.
            Brown lines and the corresponding regions indicate number of samples an computational budget during the initial MLMC warmup iteration,
            whereas green lines indicate the total accounts of both quantities at the final MLMC iteration;
            green regions indicating the differences between these two iterations.
            We observe that despite thousand of samples required on coarser levels of resolution, the required corresponding computational budget is comparable among all levels.}
  \label{f:uq100u_samples_budget}
\end{figure}

For the optimization of samples numbers $M_\ell$ on each level,
a prescribed budget of 16 million core hours was used
and the optimal number of samples was determined by \eqref{eq:M-budget}.
Estimated number of samples $M_\ell$ for the warm-up and final iterations of the OF-MLMC algorithm are depicted in \autoref{f:uq100u_samples_budget}, 
together with the resulting estimated computational budget on each level.
We observe that a significantly larger number of samples is required on the coarser levels of resolution owing to a strong reduction in level difference variances $\tilde \sigma^2_{\ell,\ell-1}$, which are also highest at the coarsest resolution levels. However, the required computational budget is comparable across all levels; such multi-level approach achieves a significant (more than two orders of magnitude) reduction in statistical error (i.e., in the variance of the statistical estimators), while at the same time keeping the deterministic error (bias) small, which is controlled by the resolution of the finest level.

In \autoref{f:comparison-MC-MLMC-OF},
a comparison of MC, MLMC, and OF-MLMC methods and estimated computational speedups is provided for a target statistical error of $1.3 \cdot 10^{-2}$.
Note, that the number of samples on the same resolution as the finest level in (OF-)MLMC and resulting computational budget required for MC simulations are estimated by
\begin{equation}
\label{eq:speedup}
M_{\text{MC}} = \left\lceil \frac{\sigma_L}{\varepsilon_{\text{OF}}} \right\rceil,
\qquad
\Work_{\text{MC}} = M_{\text{MC}} \Work_L.
\end{equation}
OF-MLMC is estimated to be more than two orders of magnitude faster than the plain MC method, and even more than three times faster than standard MLMC method without optimized control variate coefficients. The overall computational budget of OF-MLMC was only approximately 8 times larger than a single FVM solve on the finest resolution level.

\begin{table}[ht]
  \centering
  \begin{tabular}{|c|c|c|c|c|}
  \hline
  ~ & $\{M_\ell\}_{\ell = 0, \dots, L}$ & budget $\mathcal{B}$ & error $\varepsilon$ & speedup \\
  \hline
  OF-MLMC & 6400, 384, 40, 2 & 16.6 M CPU hours & $1.3\cdot 10^{-2}$ & 176.8 \\
  \hline
  MLMC & 4352, 258, 32, 3 & $\sim$ 50 M CPU hours & $1.3\cdot 10^{-2}$ & 50.6 \\
  \hline
  MC & $\sim$ 2 million & $\sim$ 2 B CPU hours & $1.3\cdot 10^{-2}$ & -\\
  \hline
  \end{tabular}
  \caption[Comparison of MC, MLMC, and OF-MLMC methods and estimated computational speedups]
            {Comparison of MC, MLMC, and OF-MLMC methods and estimated computational speedups over standard MC.
            OF-MLMC is estimated to be more than two orders of magnitude faster than plain MC, and even more than three times faster than standard MLMC.}
  \label{f:comparison-MC-MLMC-OF}
\end{table}

\subsection{Statistics for temporal quantities of interest}
\label{ss:results-temporal}

\begin{figure}[ht]
  \centering
  \includegraphics[width=0.49\textwidth]{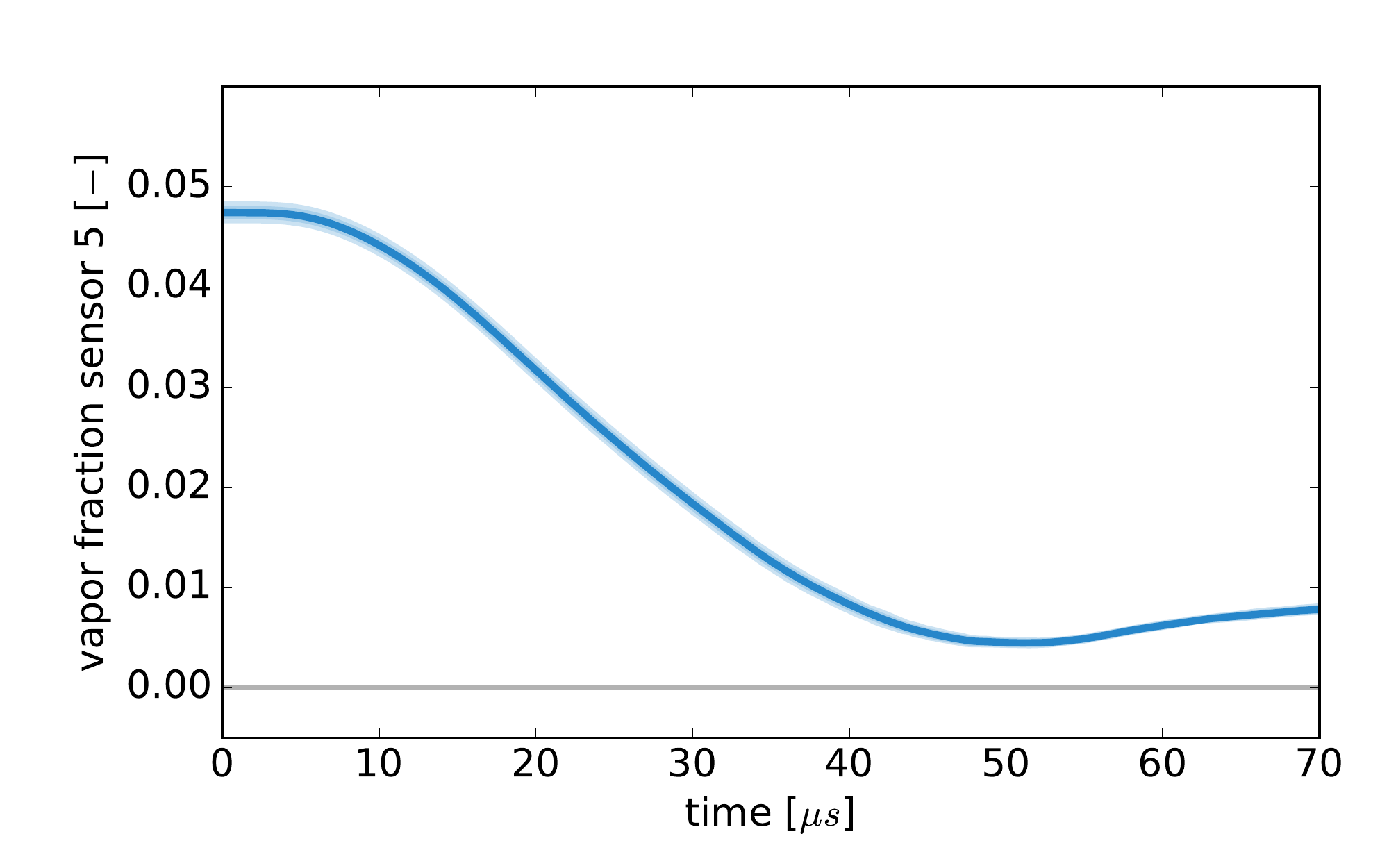}
  \includegraphics[width=0.49\textwidth]{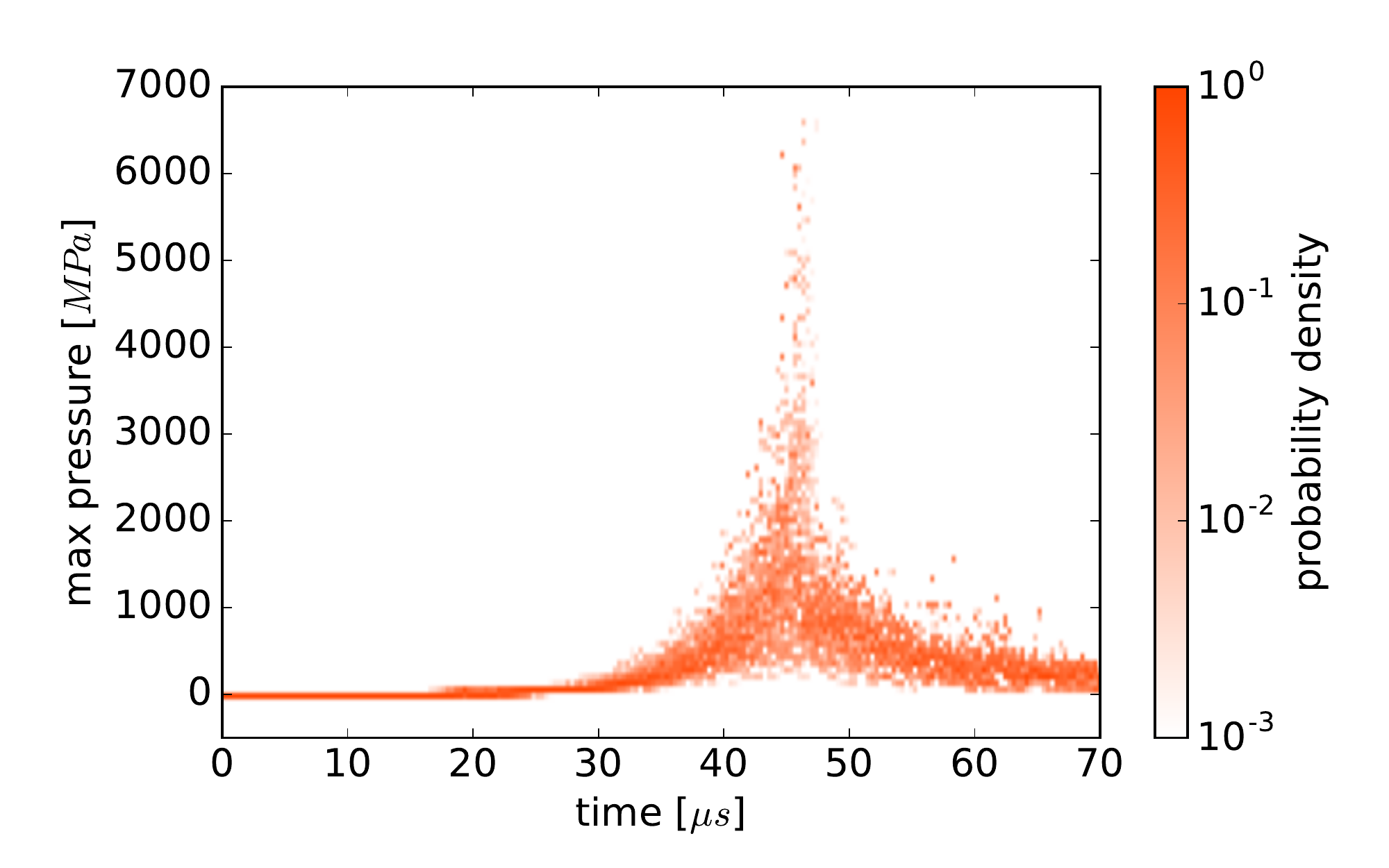}
  \caption[Uncertainties in the cloud gas volume and global peak pressure during the collapse]
            {Uncertainties in the cloud gas volume (mean values with 50\% and 90\% confidence intervals, left) and global maximum pressure (scaled empirical histogram, right) within the cloud during the collapse.
             Since all initial cloud configurations contain the same number of equally-sized cavities, very low uncertainties are observed in the evolution of the total gas volume. However, the statistical spread of the peak pressure is especially wide at its maximum value, implying a large uncertainty in the damage potential of the cavitation collapse.}
  \label{f:uqvf500lFF_mlmc_ap}
\end{figure}

\begin{figure}[ht]
  \centering
  \includegraphics[width=0.49\textwidth]{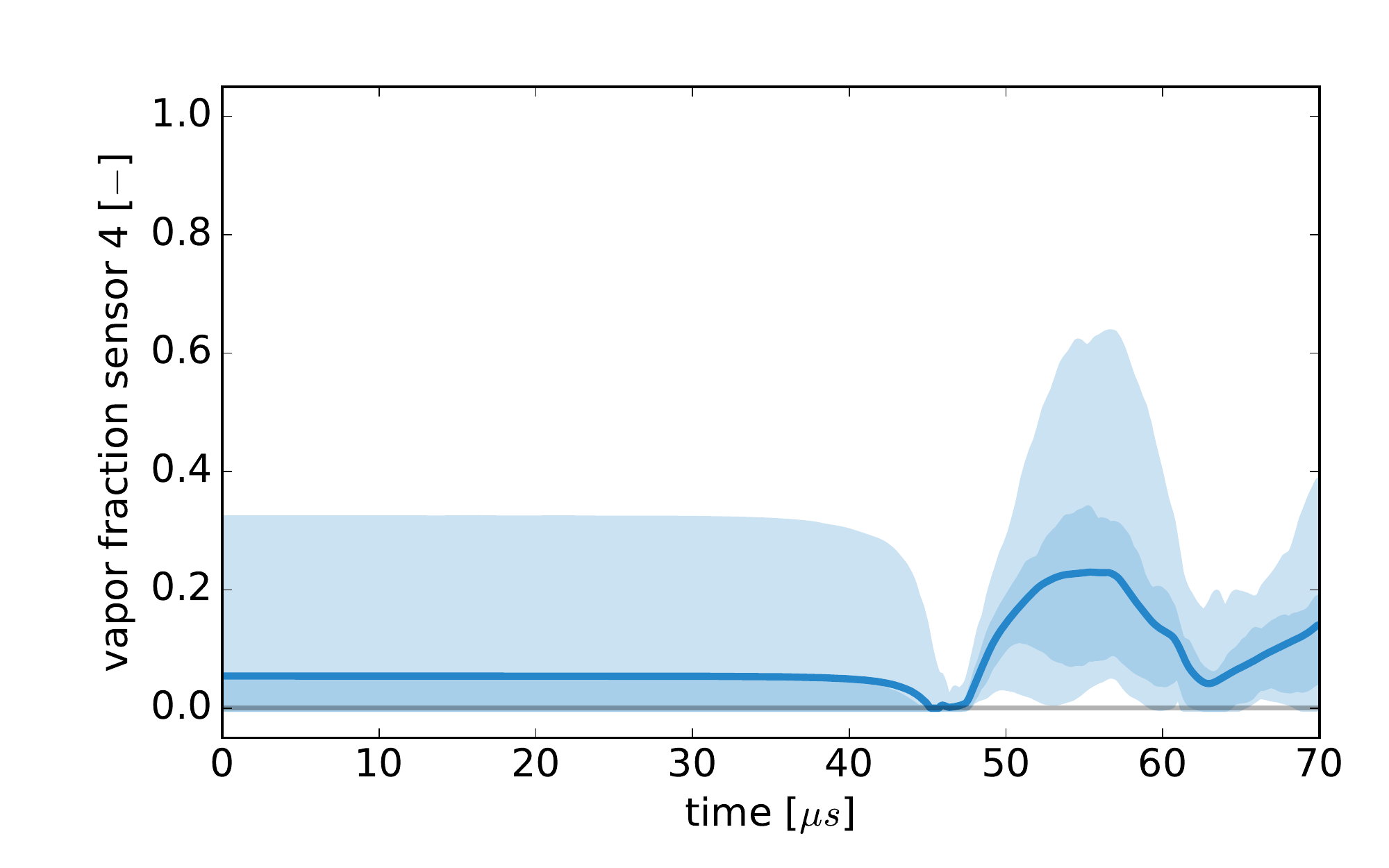}
  \includegraphics[width=0.49\textwidth]{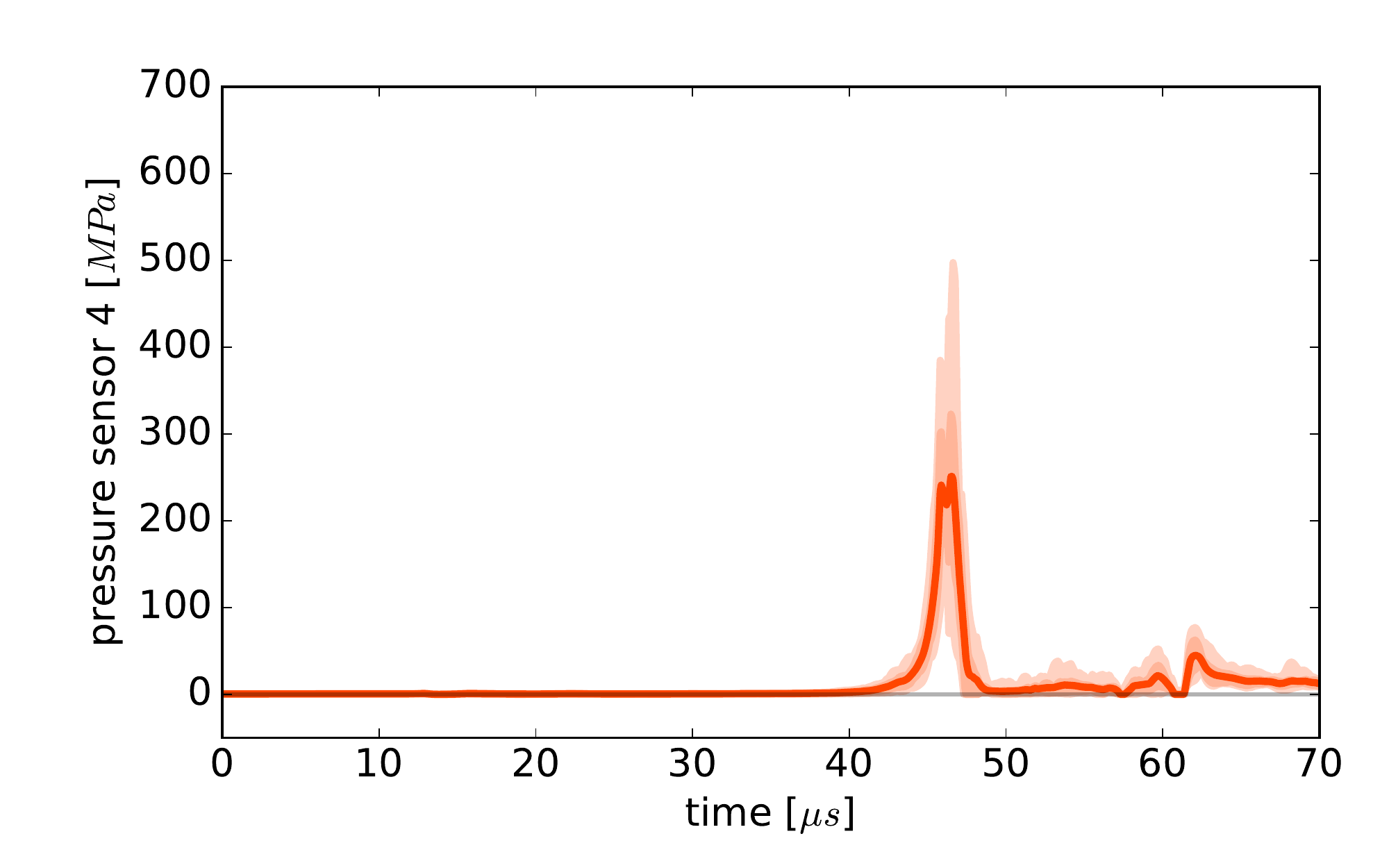}
  \caption[Uncertainties in the gas volume and pressure sensors during the collapse]
            {Uncertainties (mean values with 50\% and 90\% confidence intervals) in the gas volume (left) and pressure (right) sensor at the center of the cloud during the collapse.
             Notice, that the statistical spread of the peak sensor pressure is especially wide at its maximum value and the post-collapse increase in the gas fraction during the rebound stage.}
  \label{f:uqvf500lFF_mlmc_sensors}
\end{figure}

\begin{figure}[ht]
  \centering
  \includegraphics[width=0.49\textwidth]{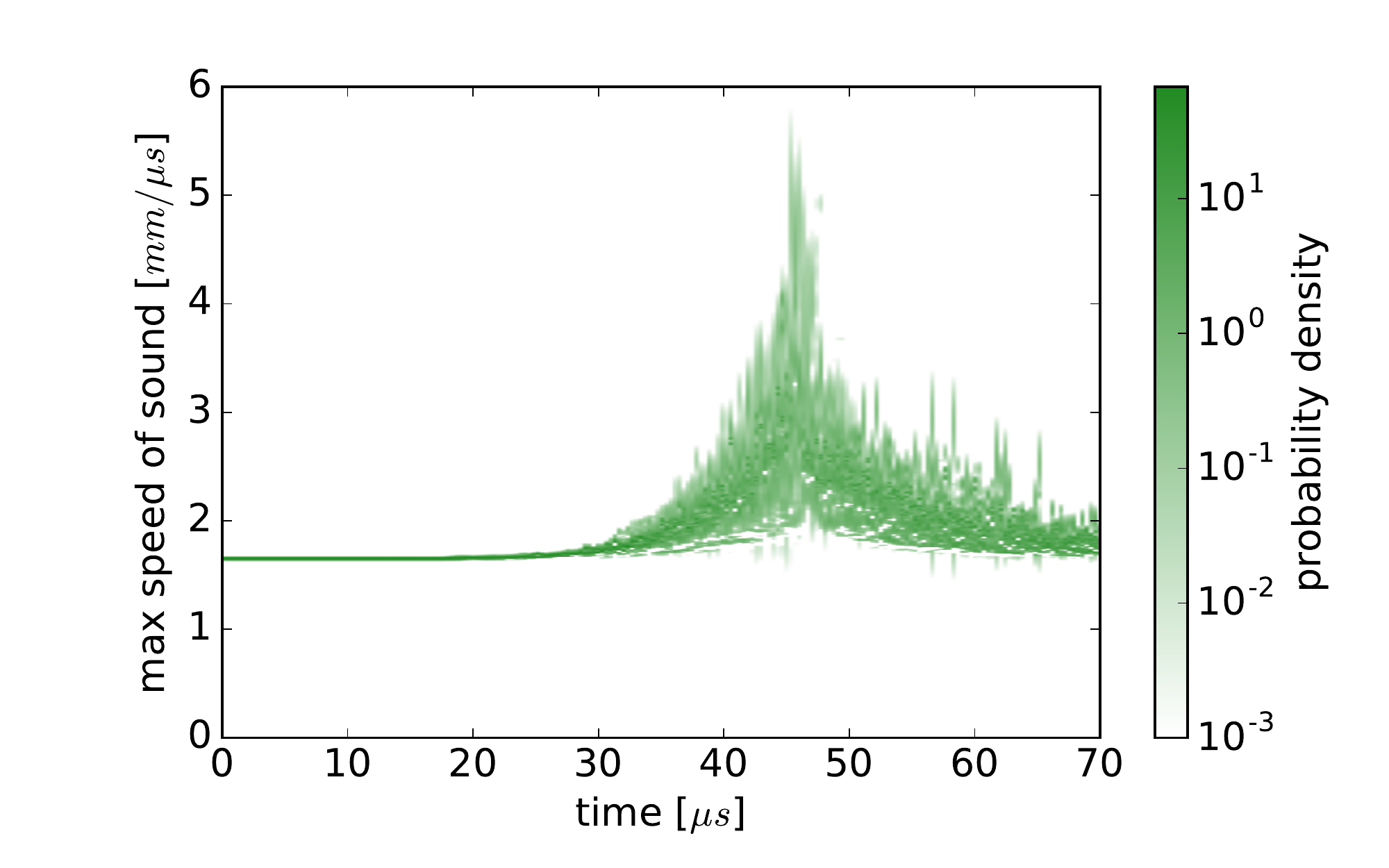}
  \includegraphics[width=0.49\textwidth]{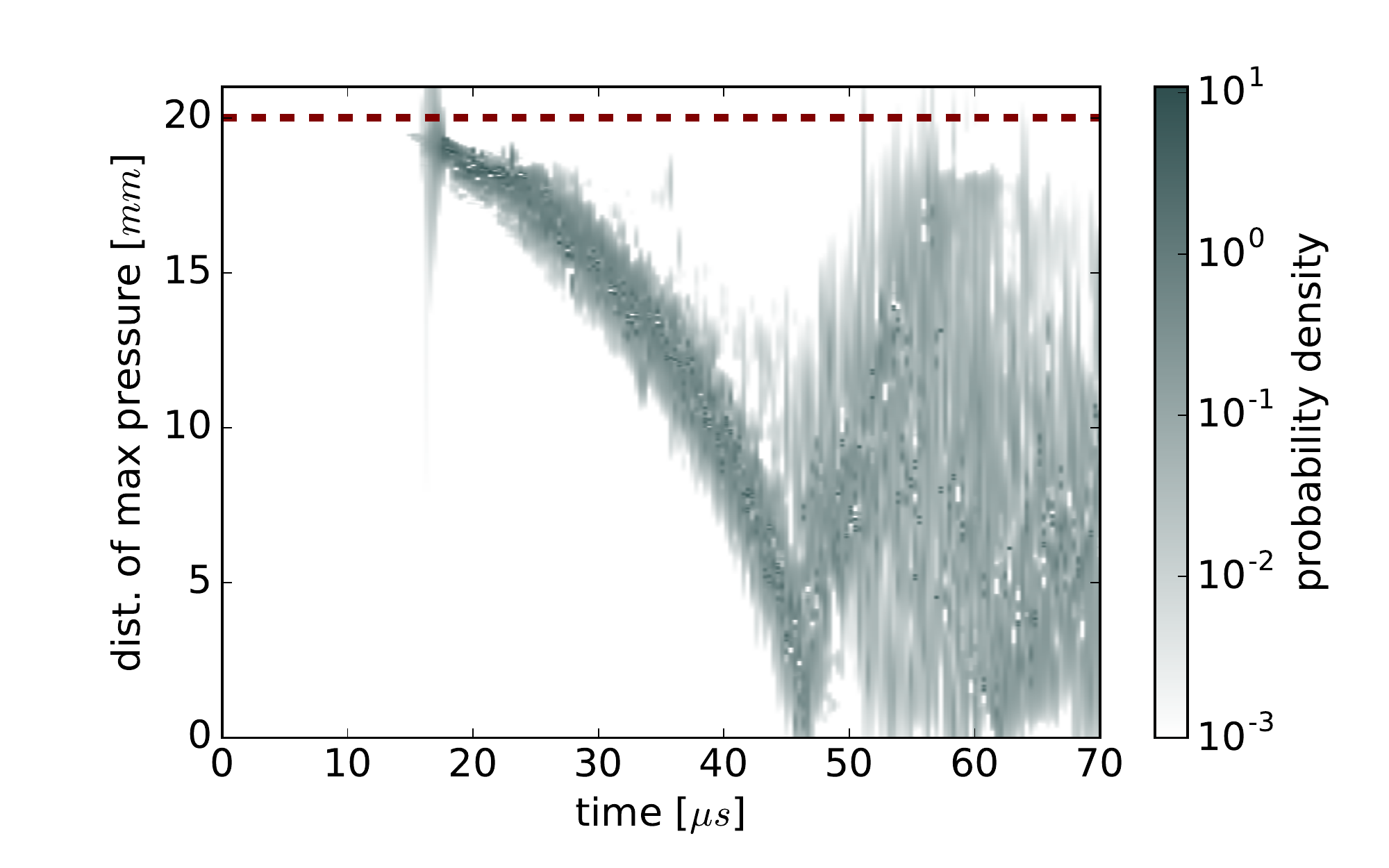}
  \caption[Uncertainties in the maximum speed of sound and peak pressure location during the collapse]
            {Uncertainties (probability density functions) in the maximum speed of sound (left) and peak pressure location (right) during the collapse.
             Dashed line indicates the initial location of the cloud surface.
             }
  \label{f:uqvf500lFF_mlmc_apd}
\end{figure}

Multi-level Monte Carlo statistical estimates are depicted in \autoref{f:uqvf500lFF_mlmc_ap}, \autoref{f:uqvf500lFF_mlmc_sensors} and \autoref{f:uqvf500lFF_mlmc_apd}.
The statistical spread of the maximum (in physical domain) pressure is especially wide at its peak (in time) value, implying a large uncertainty in the damage potential of the cavitation collapse.
To the best of our knowledge, such full Probability Density Functions (PDFs) are reported here for the first time when using the MLMC methodology for non-linear systems of conservations laws.
To obtain such estimates, level-dependent kernel density estimators were used,
with maximum bandwidth determined using a well-known ``solve-the-equation'' method \cite{SJ91,RD06}.
%
%
Such empirical PDFs are significantly more valuable in engineering,
compared to less informative mean/median and deviations/percentiles estimators,
especially for bi-modal probability distributions often encountered in such non-linear systems of conservations laws due to the presence of shocks \cite{MSS11-syscl}.

In \autoref{f:uqvf500lFF_mlmc_ap}, uncertainties in the cloud gas volume (represented by gas fraction sensor \#5, located at the center $\bc = (50\ mm, 50\ mm, 50\ mm)^\top$ with $20\ mm$ radius (hence containing the entire cloud) and global maximum pressure within the cloud are measured during the entire collapse of $70\ \mu s$.
As all initial cloud configurations contain the same number of equally-sized cavities, very low uncertainties are observed in the evolution of the total cloud gas volume. However, the statistical spread of the peak pressure is especially wide at its maximum value, indicating a strong necessity for uncertainty quantification in such complex multi-phase flows.

In \autoref{f:uqvf500lFF_mlmc_sensors}, uncertainties are measured in the gas volume fraction sensor $\alpha_\bc$ and pressure sensor $p_\bc$ at the center of the cloud as defined in \eqref{eq:sensor}.
Similarly as for the observations of peak pressure behavior, the statistical spread of the peak sensor pressure is especially wide at its maximum value. The post-collapse increase in the gas fraction indicates the presence of a  rebound stage.
During this stage, the post-collapse gas fraction consistently (in terms of confidence intervals) exceeds pre-collapse values, indicating the presence of induced outgoing rarefaction waves.

In \autoref {f:uqvf500lFF_mlmc_apd}, uncertainties in the maximum speed of sound and the peak pressure location distance from the center of the cloud are measured during the entire collapse of $70\ \mu s$.
The uncertainty in the maximum speed of sound is a direct consequence of large uncertainties in the global peak pressure.
However, on the contrary, the uncertainty in the distance of the peak pressure location from the cloud center is much smaller,
i.e., the temporal-spatial profile of the pressure wave evolution as it travels from the surface of the cloud towards the center has a much lower uncertainty (when compared to the large observed uncertainties in the global maximum pressure estimates).

\subsection{Statistics for spatial quantities of interest}
\label{ss:results-spatial}

\begin{figure}[ht]
  \centering
  \includegraphics[width=0.49\textwidth]{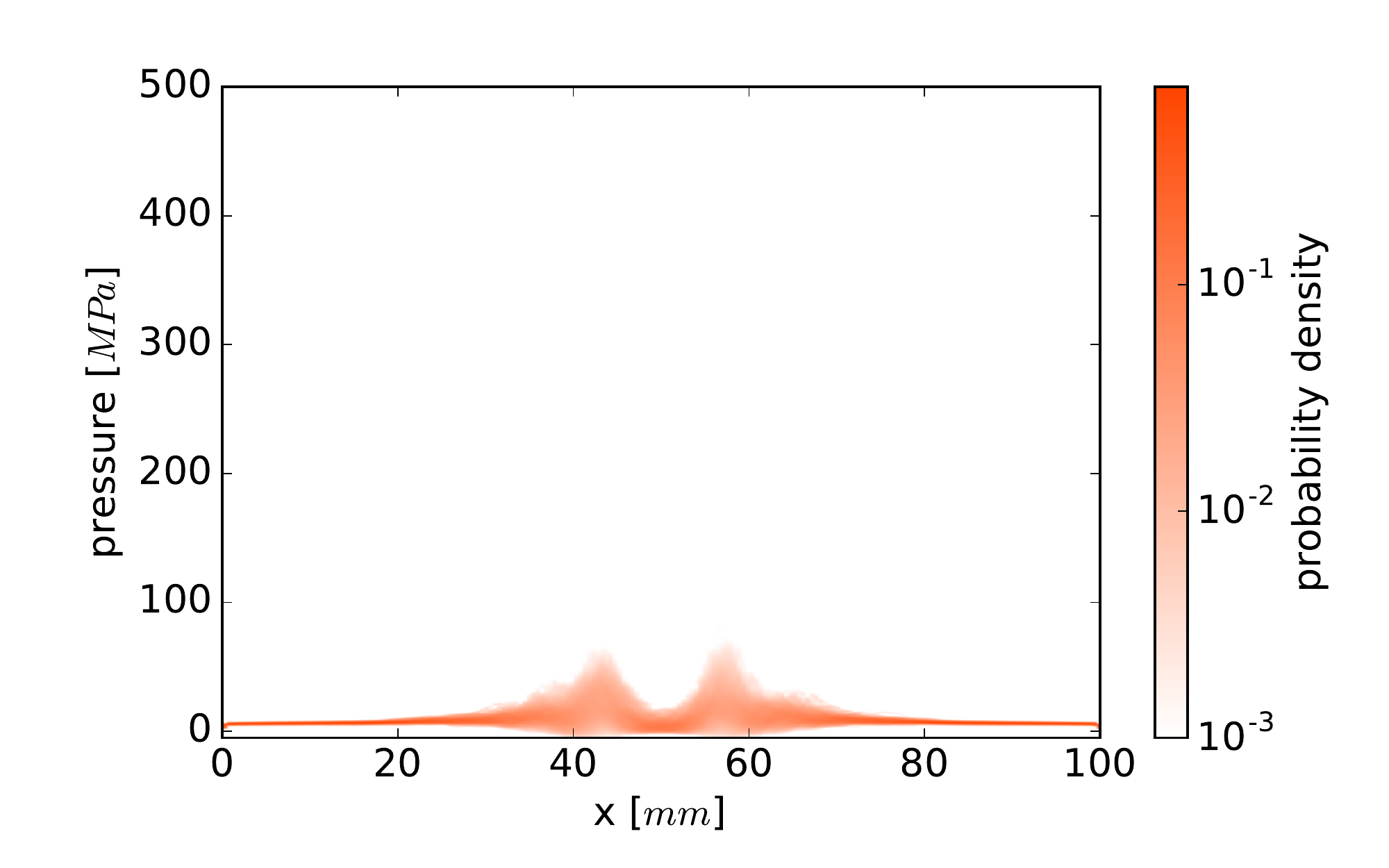}
  \includegraphics[width=0.49\textwidth]{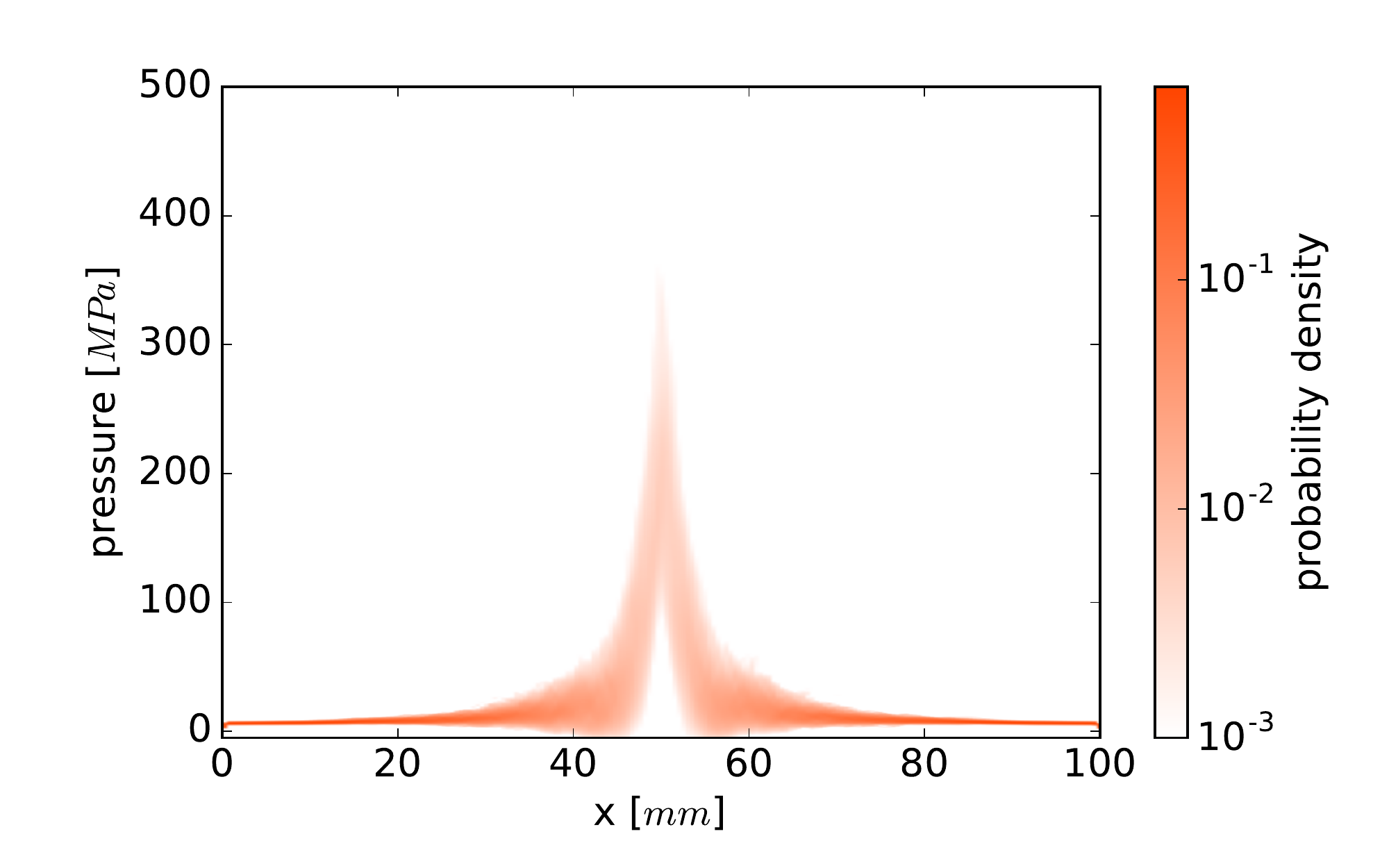}
  \caption[MLMC-estimated uncertainties of the extracted pressure line]
            {MLMC-estimated uncertainties (PDFs) of the extracted pressure line at the pre-collapse time ($t = 43\ \mu s$, left) and at the time of largest peak pressure ($\sim46\ \mu s$, right). The resulting uncertainty in encountered pressures increases significantly at the final collapse stage, with largest spreads observed near the epicenter of the cloud cavitation collapse.}
  \label{f:uqvf500lFF_mlmc_line_p}
\end{figure}

In this section, we plot the statistical estimates of QoIs extracted along one-dimensional lines
that are cut as a straight line through the center of the cloud in the three-dimensional computational domain.
We note that radial symmetry is assumed in the statistical distribution of cavities within the cloud,
and hence such one-dimensional statistical estimates through the center of the cloud are sufficient to represent the global uncertainty in the entire three-dimensional domain.
The objective of extracted one-dimensional line plots is to provide a better insight into uncertainty structures at the center of the cloud
by plotting all statistical estimates in a single plot. 
The line is cut at a specific physical simulation time, when the peak pressure is observed, and hence is slightly different for every sample.
In order to reduce volatility in global maximum pressure measurements and hence the choice of the collapse time, we smoothen the observed peak pressure measurements with a Gaussian kernel of width $0.5\ mm$ by the means of fast Fourier transform.
Statistical estimates for such extracted one-dimensional lines for pressure at different stages of the collapse process are provided in \autoref{f:uqvf500lFF_mlmc_line_p}.


The uncertainties are estimated using the MLMC methodology in the extracted pressure along the line in $x$-direction (with coordinates $y = 50\ mm$ and $z = 50\ mm$ fixed) at the pre-collapse time $t = 43\ \mu s$ and at the time of largest peak pressure, which occurs approximately at $46\ \mu s$. The time of largest peak pressure depends on the initial could configuration and hence is a random variable, varying in each statistical realization.
We observe that the resulting uncertainty in encountered pressures increases significantly at the final collapse stage, and largest spreads are observed near the epicenter of the cloud cavitation collapse, where the damage potential is the highest.

\subsection{Analysis of linear and non-linear dependencies}
\label{ss:results-dependencies}

Statistical estimates reported in the previous sections indicate that even though the initial cloud setup is very similar for all realizations,
i.e., equal count of cavity bubbles, identical cloud radius and cavity radii ranges (which resulted in very small uncertainties for the cloud volume reported in \autoref{f:uqvf500lFF_mlmc_ap}), and equal initial gas and liquid pressures, the resulting peak pressure uncertainty is very large,
as seen in \autoref{f:uqvf500lFF_mlmc_sensors} and \autoref{f:uqvf500lFF_mlmc_line_p}.

\begin{figure}[ht]
  \centering
  \includegraphics[height=170pt, trim = 90 0 220 0, clip]{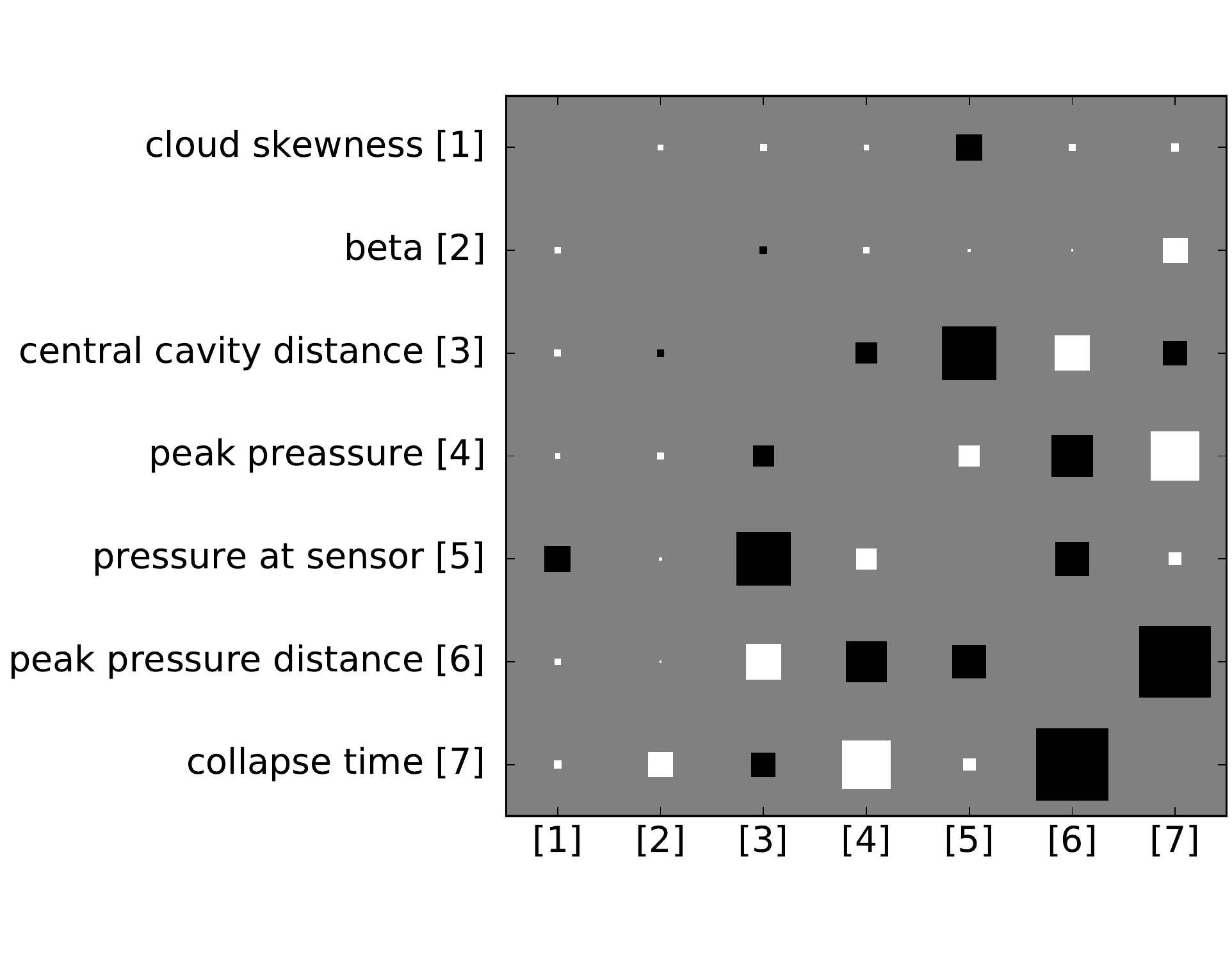}
  \includegraphics[height=170pt, trim = 282 0 220 0, clip]{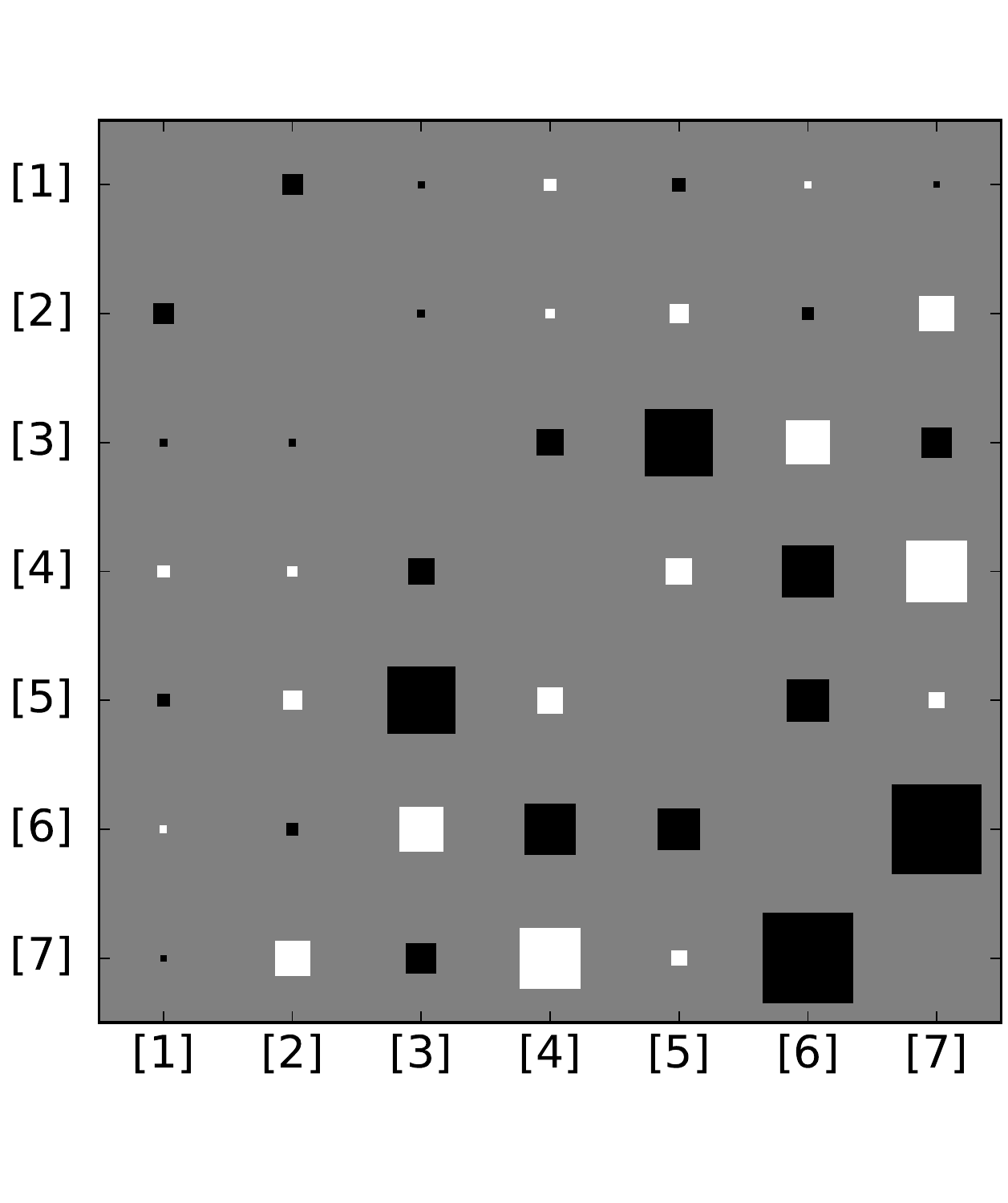}
  \caption[Hinton diagram indicating pair-wise linear statistical correlations]
            {Hinton diagram indicating pair-wise linear statistical correlations between candidate metrics and selected quantities of interest. White squares indicate positive and black squares indicate negative correlations. The size of the square indicates the magnitude of the correlation in the interval $[0,1]$.
            Cavities within the entire cloud and only in the core of the cloud are considered in the left and right plots, respectively.}
  \label{f:uqvf500lFF_mlmc_selection_correlations}
\end{figure}

Hence, it is only the actual configuration (or distribution) of the cavities inside the cloud that can have such amplifying (or attenuating)
effect on the peak pressures.
The main scope of this section is to investigate various quantities of interest that could potentially explain the cause of such non-linear effects.
The set of selected candidate metrics for the cloud configuration includes skewness (asymmetry) of the initial spatial cavity distribution inside the cloud, cloud interaction parameter $\beta$, and distance (from the center of the cloud) of the central cavity (i.e. the cavity closest to the center).
Cloud skewness is a measure of asymmetry of the cloud and is estimated by a statistical centered third moment of the distribution of cavity locations along each of the three spatial axes.
All quantities from this set of candidate metrics are tested for linear statistical correlations with the observed values of peak pressure, peak pressure distance from the center of the cloud, peak pressure at the sensor at the center of the domain, and collapse time (when largest peak pressure occurs). In addition, we have also tested for statistical correlations among QoIs themselves, such as peak pressure location and the location of the center-most cavity in the cloud.
The results are provided as a Hinton diagram in \autoref{f:uqvf500lFF_mlmc_selection_correlations}.

We observe multiple significant direct and inverse linear statistical correlations between candidate metrics and QoIs:
\begin{itemize}
	\item mild \emph{inverse} correlation between \emph{cloud skewness} and \emph{pressure sensor} readings, mainly a consequence of the central sensor placement within the cloud;
	\item strong correlation between the \emph{location of central cavity} and \emph{location of peak pressure} (w.r.t. cloud center),
	confirming prior observations in \cite{paramstudy} that peak pressures in the cloud are observed \emph{within} cavities that are near the center of the cloud;
	\item strong \emph{inverse} correlations between \emph{peak pressure location} and \emph{peak pressure magnitude}, indicating that highest peak pressures are most probable near the center of a cloud;
	\item moderate correlation between \emph{$\beta$} and \emph{collapse time},
	since large $\beta$ values can be a consequence of large gas fraction in the cloud.
\end{itemize}
%
%

Despite numerous correlations explaining the statistical spread of observed pressures,
the influence of cloud interaction parameter $\beta$
remains undetermined.
To this end, we consider cloud skewness and $\beta$ only for the \emph{core} of the cloud. We have identified the core of the cloud to be a region around the center of the cloud where uncertainties in peak pressure are largest, resulting in a spherical core with radius of $10\ mm$, based on results in \autoref{f:uqvf500lFF_mlmc_line_p}. In this case, correlations involving respective metrics such as cloud skewness and $\beta$ for the core of the cloud, are observed to be more significant:
\begin{itemize}
\item mild direct correlation between \emph{$\beta$} and \emph{pressure sensor} readings, indicating stronger collapse for clouds with higher cloud interaction parameter $\beta$ due to stronger pressure amplification;
\item mild inverse correlation between cloud skewness and \emph{$\beta$}.
\end{itemize}
Overall, such insight into statistical correlations provides a very informative description of inter-dependencies between cloud properties and observed collapse pressures, suggesting direct causal relations w.r.t.
cloud non-skewness, interaction parameter $\beta$, and proximity of the central cavity to the cloud center.

\begin{figure}[ht]
  \centering
  \includegraphics[width=0.49\textwidth]{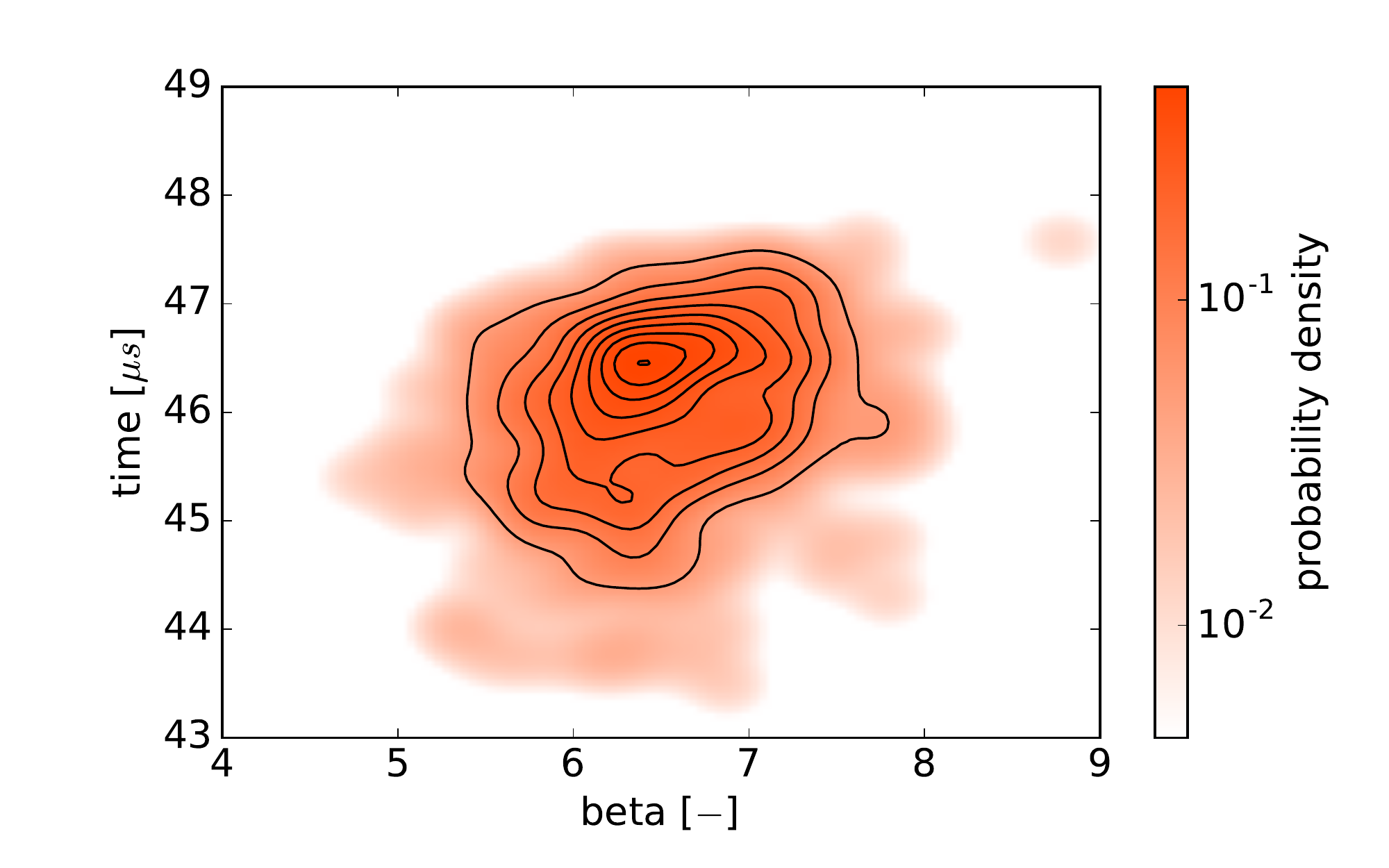}
  \includegraphics[width=0.49\textwidth]{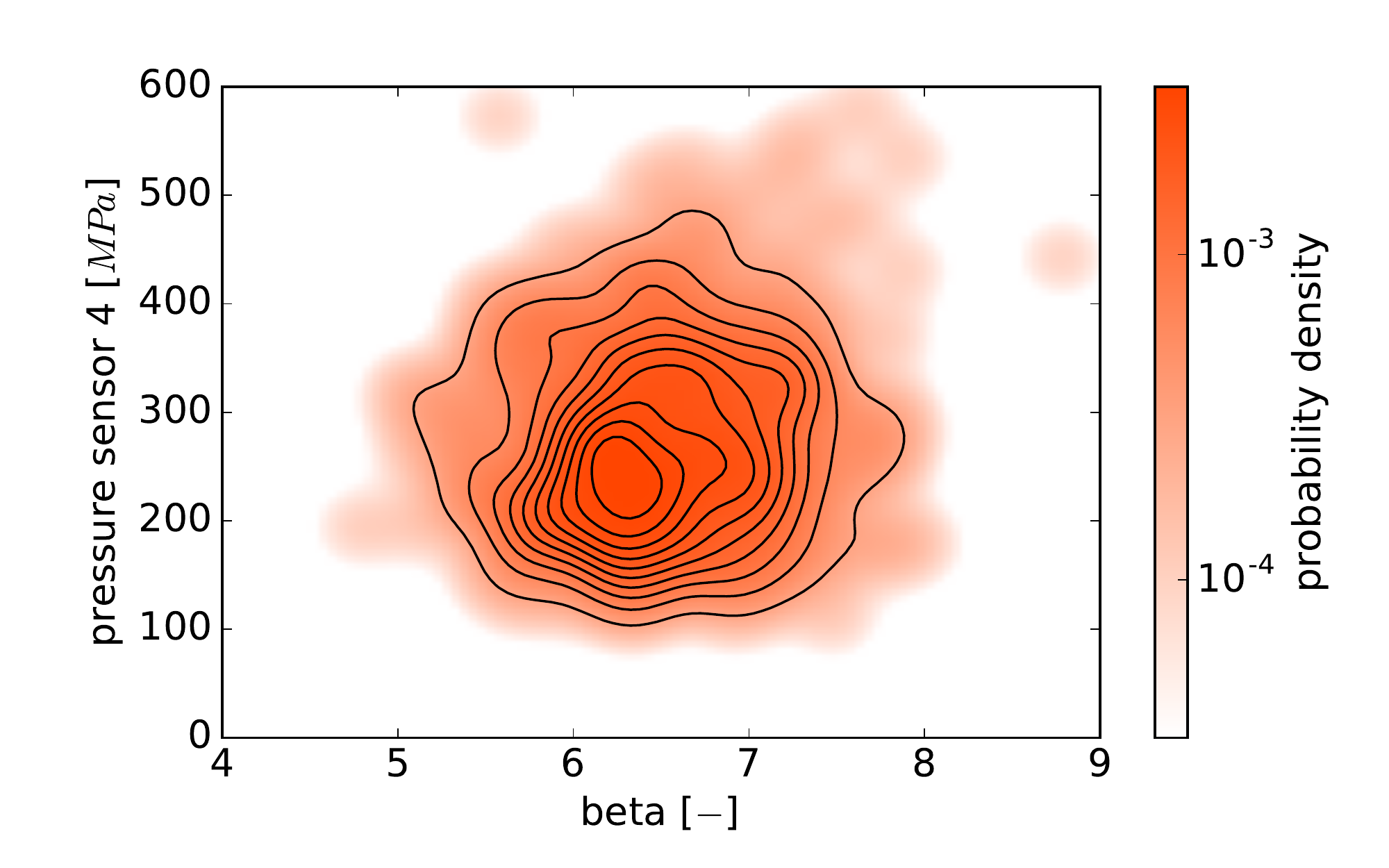}
  \caption[Joint probability distributions of cloud interaction parameter with collapse time and pressure sensor]
            {Joint PDFs of cloud interaction parameter $\beta$ with collapse time (left) and the resulting pressure sensor reading (right).
            Higher values of cloud interaction parameter $\beta$ are more likely to cause larger and more delayed collapse pressures.}
  \label{f:uqvf500lFF_mlmc_selection_kde2d-global}
\end{figure}

\begin{figure}[ht]
	\centering
	\includegraphics[width=0.49\textwidth]{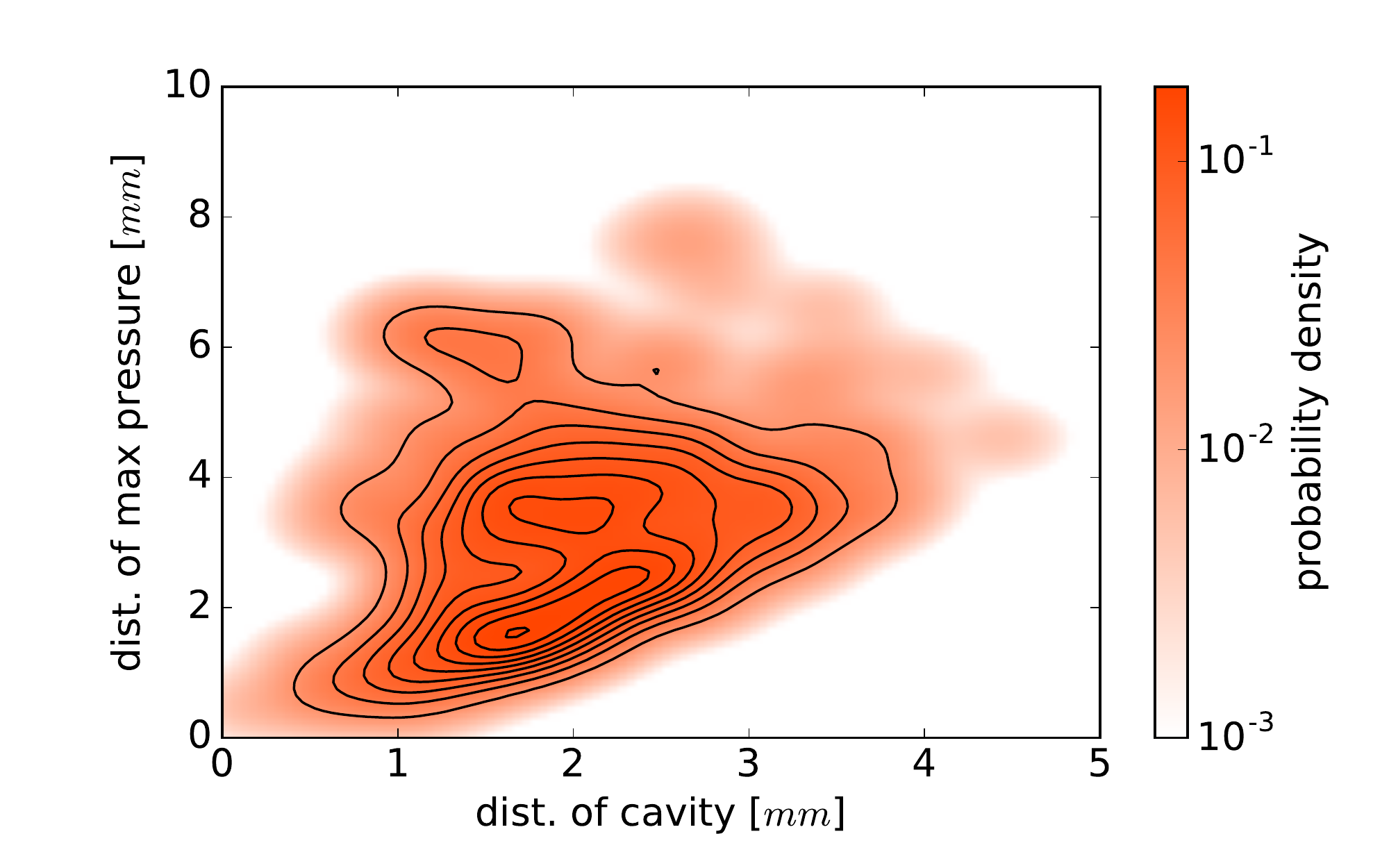}
	\includegraphics[width=0.49\textwidth]{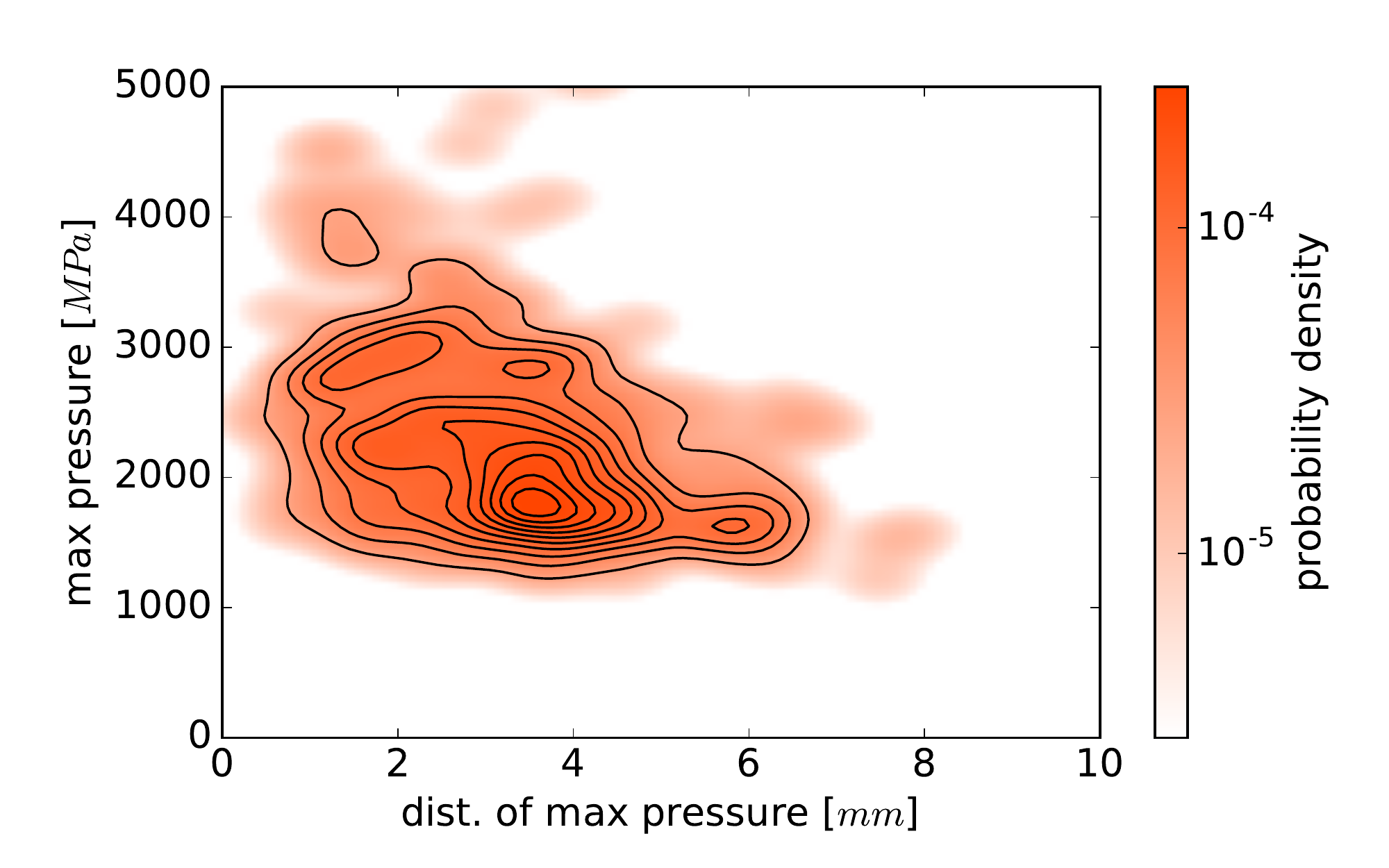}
	\caption[Joint probability distributions of location of central cavity and peak pressure magnitude with location of peak pressure]
	{Joint PDFs of location of central cavity and location of peak pressure (left), and location of peak pressure and peak pressure magnitude (right).
        The location of the central cavity bubble correlates strongly with peak pressure location, which itself exhibits strong \emph{inverse} correlations with the magnitude of the collapse pressures, explaining a wide 
        confidence intervals observed in \autoref{f:uqvf500lFF_mlmc_ap} and \autoref{f:uqvf500lFF_mlmc_line_p}.}
	\label{f:uqvf500lFF_mlmc_selection_kde2d-local}
\end{figure}

Due to a non-linear nature of the multi-phase equations and the collapse process,
non-linear dependencies among candidate metrics and QoIs could be present,
which might be inaccurately captured by the estimates of linear statistical correlations in \autoref{f:uqvf500lFF_mlmc_selection_correlations}.
We tested this hypothesis by estimating the full joint probability distribution for the pairs of significantly correlated candidate metrics and pressure behavior observations.
In \autoref{f:uqvf500lFF_mlmc_selection_kde2d-global} and
\autoref{f:uqvf500lFF_mlmc_selection_kde2d-local}, we provide results for a selected subset of tested correlation pairs, where strongest and most relevant correlations were observed.

Joint probability distributions
are consistent with linear correlation estimates obtained in \autoref{f:uqvf500lFF_mlmc_selection_correlations},
and additionally provide valuable insight into non-linear dependencies among QoIs.
Obtained results provide a good global overview of causal links between cloud structure and collapse pressures and motivate further analysis to determine the complex mechanisms governing the dynamics of such large and complex cloud cavitation phenomena. We would also like to refer to an ongoing extensive (deterministic) parameter study \cite{paramstudy} which investigates such causal links for a wider range of cloud sizes, cavity counts and
cloud interaction parameter $\beta$ values.

\section{Summary and conclusions}
\label{s:summary}

We have presented a non-intrusive multi-level Monte Carlo methodology for uncertainty quantification in multi-phase cloud cavitation collapse flows,
together with novel optimal control variate coefficients
which maintain the efficiency of the algorithm even for weak correlations among resolution levels and
delivers significant variance reduction improvements.
We have conducted numerical experiments
for
cavitating clouds containing $500$ cloud cavities, which are randomly (uniformly) distributed within the specified $20\ mm$ radius,
and the radii of the cavity follow a log-normal distribution.
The results of these numerical experiments have revealed significant uncertainties in the magnitude of the peak pressure pulse,
emphasizing the relevance of uncertainty quantification in cavitating flows.
Furthermore, statistical correlation and joint probability density function estimates have revealed potential underlying causes of this phenomenon. In particular, spatial arrangement characteristics of the cloud and its core, such as skewness, cloud interaction parameter $\beta$, and the position of the central cavity have been observed to have a significant influence on the resulting pressure amplification intensities during the collapse process.

All numerical experiments were performed by coupling an open source
PyMLMC framework with  Cubism-MPCF, a high performance peta-scale finite volume solver.
Evolution of collapsing clouds has been simulated by explicit time stepping subject to a CFL stability constraint on a hierarchy of uniform, structured spatial meshes.
Efficient multi-level Monte Carlo sampling has been observed to exhibit more than two orders of magnitude in estimated computational speedup when compared to standard Monte Carlo methods, with an additional factor 3 estimated speedup due to optimal control variate coefficients.
In the present work, we have observed the efficient scaling of the proposed hybrid OF-MLMC-FVM method
up to the entire MIRA supercomputer consisting of half a million cores. Considering that fault-tolerance mitigation mechanisms are implemented in PyMLMC and have been successfully used, we expect it to scale linearly and be suitable for the exa-scale era of numerical computing.

The proposed OF-MLMC-FVM can deal with a very large number of sources of uncertainty. In the problems presented here, 2'000 sources of uncertainty are needed to fully describe the random initial configuration of the collapsing cloud. To the
best of our knowledge, currently no other methods (particularly deterministic methods such as quasi Monte Carlo, stochastic
Galerkin, stochastic collocation, PGD, ANOVA, or stochastic FVM) are able to handle this many sources of uncertainty, in particular for non-linear problems with solutions which exhibit path-wise low regularity and possibly non-smooth dependence on random input fields.
Furthermore, the proposed methodology is well suited for  multi-level extensions of Markov Chain Monte Carlo methods for Bayesian inversion \cite{HSS14,MCMLMC}.

The  present  multi-level methodology is a powerful general purpose technique for quantifying uncertainty in complex flows governed by hyperbolic systems of non-linear conservation laws such as cloud cavitation flow problems.

\appendix

\section{Discussion}
\label{a:discussion}

Here, we  wish to comment on possible shortcomings
in the progress of the adaptive OF-MLMC algorithm described in \autoref{sss:ocv-mlmc-adaptive}.
These remarks are meant to assist the application of  OF-MLMC  
and are not directly relevant for the numerical results presented in \autoref{s:results}.

\subsection{Number of ``warm-up'' samples on the finest resolution level}
\label{r:ML=1}
%
Notice, that in order to have \emph{empirical} estimates of $\sigma^2_\ell$, $\sigma^2_{\ell,\ell-1}$, and $\tilde\sigma^2_\ell$,
at least \emph{two} samples would be required on each level $\ell = 0, \dots, L$.
Enforcing $M_L \geq 2$ might be very inefficient in the cases when only \emph{one} sample is actually needed,
since in presently considered applications the most computationally expensive samples are actually at the finest level $\ell = L$.
To avoid this, initially only \emph{one} ``warm-up'' sample on the finest mesh level $\ell = L$ could be computed,
i.e.,~$M_L = 1$.
For subsequent optimization steps of each $M_\ell$, variance of level difference $\sigma^2_{\ell,\ell-1}$ for level $\ell = L$
is inferred from available measurements on lower resolution levels
using Bayesian inference, as described in step (3) of the OF-MLMC algorithm.
If more than one sample is actually required on the finest resolution level,
optimization step (7) of the adaptive OF-MLMC algorithm above will adjust $M_L$ to the correct optimal value and additional empirical estimate would be available for even more accurate inference.

\subsection{Iterative control of indicator errors}
\label{r:cmlmc}
%
Final OF-MLMC error $\varepsilon_{\text{OF}}$ could be underestimated by $\hat\varepsilon_{\text{OF}}$ and be actually \emph{larger} than the prescribed tolerance $\tau$,
since we only ensure the \emph{estimated} total error $\hat\varepsilon_{\text{OF}} \leq \tau$.
Since $\hat\varepsilon_{\text{OF}}$ is based on the \emph{randomized} statistical estimators,
it is also \emph{random} and has a spread around its mean $\IE[\hat\varepsilon_{\text{OF}}] = \varepsilon_{\text{OF}}$.
We note, that probability $\IP[\varepsilon_{\text{OF}} > \hat\varepsilon_{\text{OF}}]$ of the resulting MLMC error $\varepsilon_{\text{OF}}$ being \emph{larger} than the estimated error $\hat\varepsilon_{\text{OF}}$ (which is forced to be smaller than the prescribed tolerance $\tau$)
can be reduced by sufficiently \emph{increasing} the number of samples $\hat M_\ell$
according to the estimates of \emph{fourth centered moment (kurtosis)} $\kappa^2_\ell$ of the level differences.
Such estimates would provide the \emph{variance estimates
$\hat\kappa^2_\ell \approx \IE[ (\hat\sigma^2_\ell)^2 - \IE[\hat\sigma^2_\ell]^2 ]$
of the empirical variance estimators $\hat \sigma^2_\ell$}.
Then, increasing $\hat M_\ell$ by several standard deviations $\hat\kappa_\ell$ of $\hat\sigma^2_\ell$,
the required percentile level $\alpha$ of confidence interval $[\alpha/2, 1 - \alpha/2]$ of $\hat \varepsilon_{\text{OF}}$
can be reduced below the prescribed tolerance $\tau$,
this way \emph{ensuring} the required high probability $\IP[\varepsilon \leq \tau] \geq 1 - \alpha$
of true MLMC error $\varepsilon$ \emph{not exceeding} the prescribed tolerance $\tau$.
A continuation MLMC method incorporating similar techniques for estimation and control of error confidence intervals was already proposed in \cite{Tempone}.
However, updating estimates after \emph{each} computed sample could be very inefficient for large HPC application, since such incremental techniques require heavy synchronization and would make efficient load balancing on distributed many-core systems very challenging.

\subsection{Sample ``recycling''}
\label{r:recycling}
%
Alternative optimal coefficients for each level in MLMC estimator was suggested in \cite{Gunzburger,Peherstorfer2016}, where a multi-fidelity Monte Carlo method is described. There, some statistical realizations (samples) are \emph{re-used}, i.e., the same result is used in both estimates $E_{M_{\ell}}[\alpha_\ell q_\ell]$ and $-E_{M_{\ell+1}}[\alpha_\ell q_\ell]$, each contributing to a separate difference in the telescoping MLMC estimator \eqref{eq:ocv-mlmc}.
Such ``recycling'' strategy requires less sampling, however, error analysis complexity is highly increased due to additional statistical dependencies, absent in OF-MLMC method.
On the other hand, for ``recycled'' sampling, the resulting error minimization problem is separable in terms of coefficients $\alpha_\ell$ and number of samples $M_\ell$, and hence no linear system \eqref{eq:alpha-matvec} needs to be solved \cite{Gunzburger}. The linear system \eqref{eq:alpha-matvec} is, however, very small, sparse, and needs to be solved only a few times, hence is not a bottleneck of this algorithm.

\subsection{Truly optimal number of MC samples on each resolution level}
\label{r:discrete}
%
A discrete optimization problem could be formulated, avoiding round-off operations in \eqref{eq:M-tol} or \eqref{eq:M-budget},
and providing truly optimal \emph{integer} $M_\ell$, as suggested in \cite{Stefan3}.
We note, however, that such round-offs do not influence the efficiency of the method on coarser levels, where the number of samples is large. Furthermore, round-off inefficiencies are most probably over-shadowed by the used approximate estimators for $\tilde\sigma^2_\ell$.

\section*{Acknowledgments}
Authors acknowledge support from the following organizations:
Innovative and Novel Computational Impact on Theory and Experiment (INCITE) program, for awarding computer time under the project CloudPredict;
Argonne Leadership Computing Facility,
which is a DOE Office of Science User Facility supported under Contract DE-AC02-06CH11357, for providing access and support for MIRA, CETUS, and COOLEY systems.
Partnership for Advanced Computing in Europe (PRACE) projects PRA091 and Pra09\_2376, together with J\"ulich and CINECA Supercomputing Centers;
Swiss National Supercomputing Center (CSCS) for computational resources grant under project ID s500.
{J{\v S} would like to thank Stephen Wu for his contribution and fruitful discussions on the optimal variance reduction techniques in the multi-level Monte Carlo estimators,
and Timothy J. Barth for hosting him as a scientific visitor at NASA Ames Research Center (California, United States) and for collaborations on robust kernel density estimators.

\bibliographystyle{siamplain}
\bibliography{refsFabian,refsUrsula,refsINCITE,refsJonas,mendeley-static}
\end{document}